\documentclass[manuscript,nonacm]{acmart}

\usepackage{booktabs}
\usepackage{graphicx}
\usepackage{xcolor}

\setcopyright{none}
\providecommand{\needcite}[1]{}
\providecommand{\pending}[1]{}

\begin{document}

\title{Balancing Evidence and Interpretation: Historical Grounding Ratio as a Design Parameter for AI-Generated Urban Storytelling}

\author{Fuyang Zhang}
\affiliation{%
  \institution{Nanjing University\\School of Architecture and Urban Planning}
  \city{Nanjing}
  \country{China}
}
\email{zhangfuyang55@gmail.com}

\author{Maurice Benayoun}
\affiliation{%
  \institution{Head of Art and Architecture Intelligence Lab\\School of Art, Nanjing University}
  \city{Nanjing}
  \country{China}
}
\email{m.benayoun@nju.edu.cn}

\begin{abstract}
Location-aware generative systems can now select historical archives and real-time contextual information based on a user's surroundings to automatically generate narratives for urban heritage walks. Yet when multiple sources jointly inform generation, existing systems provide neither a clear representation of how much content from each source actually appears in the output nor an operational means of measuring it. We introduce the Historical Grounding Ratio (HGR), defined as the proportion of claim-bearing information units in a generated narrative that are supported by historical archives. HGR turns the realized share of historical evidence in a narrative into a directly measurable design parameter. In GeoDrama, a mobile narrative system, we created three conditions that used a common retrieval procedure and comparable evidence-bundle sizes while varying the allocation of information from different sources during generation. We evaluated how changes in HGR affected narrative experience through a within-subject walking study with 18 participants. Increasing HGR significantly strengthened the perceived relevance between narrative content and the specific location. However, historical understanding, integration with the visible scene, appropriateness of the amount of information, and intention to explore further did not increase monotonically with HGR; all four measures were highest in the intermediate, balanced condition. These findings show that designing location-aware generative interfaces involves not only retrieving relevant material but also determining how information from different sources composes the final output. HGR offers an operational measure for comparing information-allocation strategies and their experiential consequences.
\end{abstract}



\maketitle

\begin{figure*}[t]
  \centering
  \includegraphics[width=\textwidth]{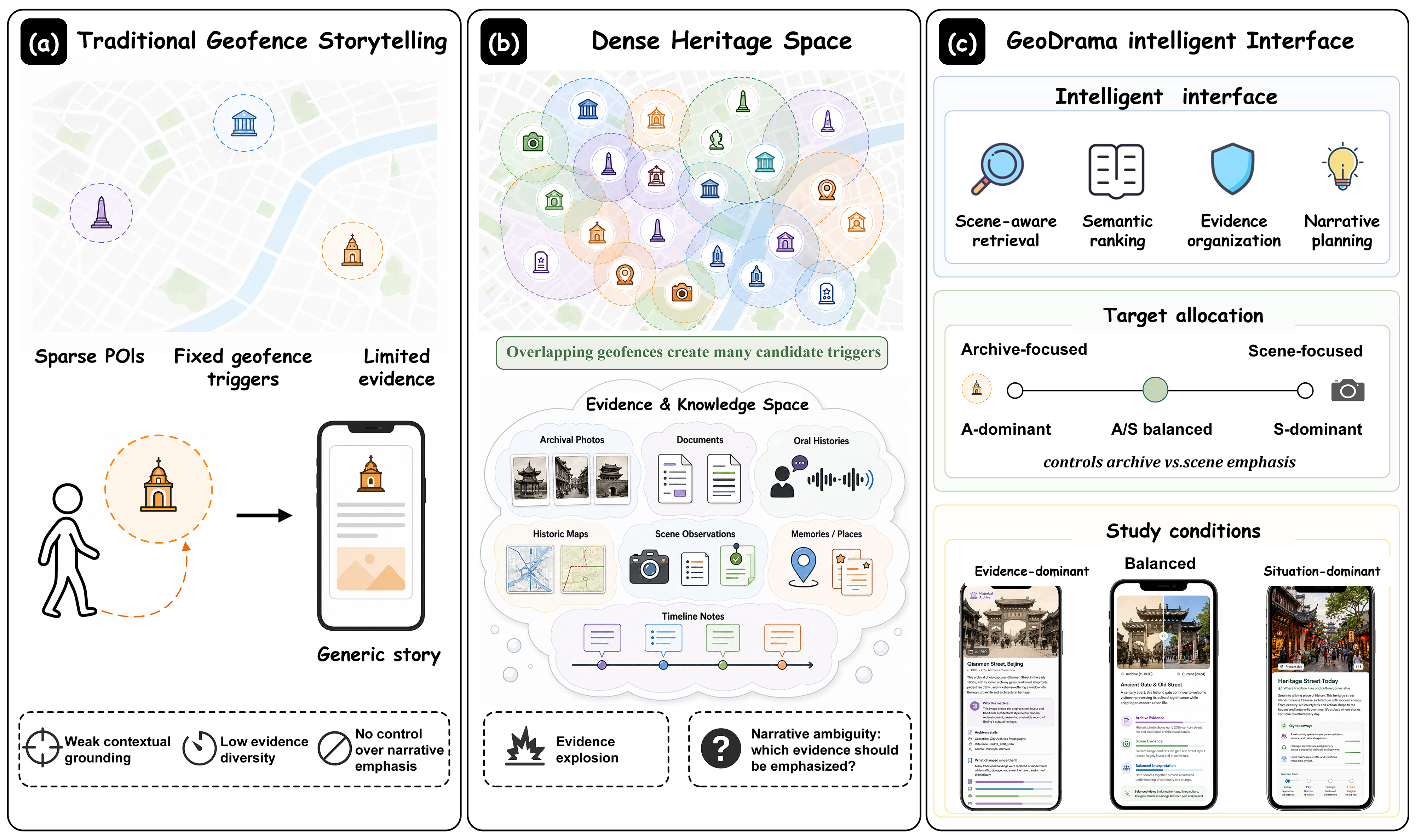}
  \caption{Design motivation and study framing for GeoDrama.
  \textbf{(a)} Fixed-geofence storytelling provides sparse triggers and limited evidence.
  \textbf{(b)} Dense heritage spaces introduce overlapping place candidates and heterogeneous historical sources.
  \textbf{(c)} GeoDrama operationalizes evidence allocation through HGR and compares situation-dominant, balanced, and evidence-dominant narrative configurations.}
  \Description{Three-panel overview of GeoDrama's design motivation and study framing. Panel a shows the sparse triggers and limited evidence of fixed-geofence storytelling. Panel b shows overlapping place candidates and heterogeneous historical sources in a dense heritage space. Panel c shows the target evidence-allocation strategy and the three study conditions.}
  \label{fig:teaser}
\end{figure*}

\section{Introduction}
\label{sec:intro}

Early location-aware mobile systems typically triggered place-bound audio, text, or other pre-authored media when a user reached a point of interest (POI) or entered a predefined geofence~\cite{bederson1995audio,priandani2017malang}. As context-aware computing developed, systems began to incorporate user preferences, interaction histories, movement trajectories, time, and environmental cues captured by cameras to determine what information to provide and when~\cite{abowd1997cyberguide,cheverst2000guide,stenton2007mediascapes,park2024comfortable,kleftodimos2023locationar}. Generative AI extends this task from triggering and selecting candidate content to composing new content at runtime, allowing systems to generate narratives from location, the live scene, user state, and external knowledge~\cite{ho2025visitor,dang2025authoring,cai2025aiget}. When multiple sources jointly inform generation, a system must decide not only which materials enter the generation process, but also how those materials compose the final output.

Existing approaches primarily manage external information through retrieval relevance and generation attribution: the former determines which candidate materials are relevant to the current location or object, whereas the latter checks whether generated claims are supported by retrieved sources~\cite{zhao2024dense,rashkin2023measuring,gao2023citations,liu2023verifiability,zhang2024finegrained,xia2025ground}. Yet even when both historical archives and the live scene are relevant to the current location, and the generated content is adequately supported, the system must still allocate a limited output budget across these sources. A larger share of historical evidence may convey more concrete facts, but it may leave less room for content that connects the narrative to the surrounding environment and supports comprehension while walking. A smaller share may strengthen the connection to the immediate situation while reducing the historical information users actually encounter. Therefore, \emph{how to allocate the share of content from different sources in the final output} constitutes an intelligent-interface design problem distinct from retrieval relevance and factual attribution. We refer to this problem as \emph{evidence allocation}. Evidence allocation directly determines what an interface ultimately presents to users and may shape how they understand the relationship among place, history, and the current environment. Treating it as an adjustable design dimension also provides a basis for future context-aware generative interfaces to adapt content presentation to place, task, and user state~\cite{vazquezalvarez2011auditory,lanir2013the}.

To empirically examine how different evidence allocations affect user experience, this design problem must first be operationalized as a variable that can be measured from the actual output. We therefore introduce the Historical Grounding Ratio (HGR). We segment each delivered narrative into information units whose sources can be independently assessed and classify them as archive-supported (A), scene-anchored (S), or interpretive/bridging (I). HGR is the realized proportion of archive-supported units among all A/S/I units. It measures the composition of the content users actually receive after generation, rather than the target ratio specified in a prompt; it is also distinct from whether retrieved materials are relevant or generated claims are correct. HGR does not prescribe an optimal allocation. Instead, it provides a common scale for comparing allocation strategies, testing their experiential consequences, and identifying trade-offs among design goals.

Figure~\ref{fig:teaser} summarizes this motivation and study framing: fixed-geofence storytelling provides sparse triggers and limited evidence, whereas dense heritage spaces introduce overlapping place candidates and heterogeneous historical sources. GeoDrama operationalizes this evidence-allocation problem through HGR and compares situation-dominant, balanced, and evidence-dominant narrative configurations.

Building on HGR, we developed GeoDrama, a mobile narrative system designed as a research apparatus for manipulating and evaluating evidence allocation. GeoDrama first uses GPS location to retrieve historical spatial polygons that contain the user's current position and the archival records associated with them. It then reranks candidate materials using a scene description generated from a synchronized camera feed, allowing the live scene to inform both evidence selection and subsequent narrative generation. Within this pipeline, we implemented three conditions: situation-dominant, balanced, and evidence-dominant. The conditions used the same retrieval mechanism, language model, generation parameters, and narrative structure; they also kept the amount of candidate evidence and the output information capacity within comparable ranges. Their primary difference was the quota of information units assigned to historical material and situated information in the final narrative. Eighteen participants completed a counterbalanced within-subject walking study along a fixed route through a historically dense urban area. Each participant experienced all three conditions, yielding 54 walking segments. After generation, we measured the realized HGR of each segment and examined how different evidence allocations affected perceived narrative--place relevance, historical understanding, integration with the visible scene, appropriateness of the amount of information, and intention to explore further.

The three conditions produced clearly distinct realized levels of historical grounding, demonstrating that GeoDrama could effectively manipulate evidence allocation in the final narratives. As HGR increased, participants were more likely to perceive the narrative as relevant to their specific location. However, historical understanding, integration with the visible scene, appropriateness of the amount of information, and intention to explore further did not improve in parallel; descriptively, all four measures were highest in the balanced condition. These findings show that historical grounding affects experiential goals differently: increasing archive-supported content can strengthen narrative--place relevance, but does not uniformly improve the overall walking experience. HGR is therefore better treated as a design parameter to be adjusted according to a particular experiential goal than as a generation-quality metric to be universally maximized.

This paper makes three contributions:

\begin{itemize}
  \item \textbf{A method for operationalizing evidence allocation as a measurable design variable.}
  HGR represents historical grounding as the realized proportion of archive-supported information units in a delivered narrative. By distinguishing the target grounding specified in a prompt from the grounding users actually receive, it enables source composition to be measured and compared at the information-unit level.

  \item \textbf{A situated narrative system for controlling and auditing historical grounding.}

  GeoDrama integrates location-constrained archival retrieval, live scene sensing, candidate reranking, and controlled narrative generation. It enables evidence allocation to be manipulated while keeping the retrieval procedure and input capacity comparable across conditions, and makes the process from retrieved evidence to delivered narrative traceable.

  \item \textbf{A field study showing that historical grounding affects walking experiences in differentiated ways.}
  Our within-subject study with 18 participants shows that increasing HGR significantly strengthens narrative--place relevance, but that this benefit does not extend in parallel to historical understanding, integration with the visible scene, appropriateness of the amount of information, or intention to explore further. The result positions HGR not as a quality metric for which higher is always better, but as an interface-policy parameter whose setting should depend on the design goal.
\end{itemize}

\section{Related Work}
\label{sec:related}

\subsection{Location-Aware and Context-Aware Storytelling}
\label{sec:related:location}

Delivering content to mobile users based on their location is one of the earliest interaction paradigms in mobile HCI. Early systems typically selected pre-authored content according to a user's location and presented it upon arrival at the corresponding place~\cite{bederson1995audio,abowd1997cyberguide,stenton2007mediascapes}. As mobile devices and positioning technologies developed, geofencing became a common implementation: entering a designated area could trigger audio or text associated with that location~\cite{priandani2017malang}. Subsequent context-aware research gradually expanded the range of information used for content selection. Beyond location, personal interests and environmental conditions can jointly inform recommendations~\cite{cheverst2000guide}; route comfort and scenic attractiveness can shape outdoor narrative experiences~\cite{park2024comfortable}; and cameras and augmented reality can connect objects in the immediate environment with supplementary information~\cite{kleftodimos2023locationar}. Location, route, user state, and the surrounding environment have thus become contextual signals that mobile interfaces can continuously sense and use to select content.

Research in mobile settings has also long examined how content should be presented as a walk unfolds. Proactive delivery in location-aware guides can affect users' sense of autonomy and redistribute their attention between the physical environment and the mobile device~\cite{lanir2013the}. Because pedestrians need visual attention for both the path and their surroundings, audio is often more suitable as the primary information medium than sustained screen viewing~\cite{vazquezalvarez2011auditory}. To support continuous mobile experiences, prior work has further examined the pacing of content, the relationship between participants and their environment, and how narratives unfold along a route~\cite{rogers2004ambient,reeves2005designing,dionisio2010the}. Recent work has begun to extend such designs from individual locations to continuous walking routes~\cite{walkingtalkingstick2023,smartwalkcoach2026}.

The spatial scale at which content is triggered also affects narrative continuity. Geofences centered on individual points of interest can leave long gaps in areas where such points are sparse; prior work has therefore used hierarchical geofences and continuous story sequences to cover the course of a walk~\cite{sasaki2024geofencing}. The degree of coupling between a narrative and the user's current location also shapes experience. Experiments have shown that locality affects immersion and mental imagery, although its effects do not appear uniformly across all experiential dimensions~\cite{karapanos2012locality}. Together, this literature has extensively examined what contextual information a location can provide, which content is relevant to the current user, and when and how that content should be presented.

Most narratives in these systems are authored before use. At runtime, a system may select different content based on location and other contextual signals, or vary the triggering range, presentation timing, and interaction mode, but the information contained within each narrative segment is typically determined during content production. Generative interfaces change this process. Historical materials, the live scene, and other contextual information can directly participate in the generation of new content at runtime. Consequently, the information that constitutes the final narrative is no longer a predetermined content property, but an outcome formed during generation.

\subsection{Generative and Situated Intelligent Interfaces}
\label{sec:related:generative}

Generative models allow context-aware interfaces to compose new content on demand from the current environment, rather than merely selecting from prepared materials~\cite{park2023generative,nichols2020collaborative,li2023locationaware}. This shift is already visible in travel, guided tours, wearable devices, and real-world assistance. PANDALens combines user behavior and environmental information during travel to generate content relevant to the user's current experience~\cite{cai2024pandalens}. Story-Driven uses the route, current location, environmental conditions, and journey progress to continuously adapt an automatically generated story, and evaluates this approach in a real mobile setting~\cite{belz2024storydriven}. In exhibition guidance, real-time visual sensing, knowledge-graph retrieval, and multimodal models can also jointly contribute to the generation of spoken content~\cite{ho2025visitor}. These studies demonstrate that real-world contextual information can now directly participate in content generation rather than merely determining whether existing content is triggered.

The context used in generation is no longer limited to a single location or visual cue. Wearable systems can combine real-time visual and auditory information, interaction histories, and progressively constructed user profiles to generate continuous responses~\cite{xu2024companions}. Smartwatches can use sensor-recognized task steps to provide different answers to the same user question at different stages of an activity~\cite{arakawa2024prismqa}. Satori incorporates the surrounding environment, nearby objects, behavioral history, task goals, and user state into proactive AR assistance~\cite{li2025satori}, while SocialMind further combines verbal and nonverbal behavior, social context, and user characteristics to generate in-situ interaction suggestions~\cite{yang2025socialmind}. Gaze, environmental analysis, and user information have also been used to proactively select and generate content worth presenting in the current scene~\cite{cai2025aiget}, while camera feeds combined with multimodal models can serve as user-configurable interfaces for sensing the physical world~\cite{liu2025gensors}.

Representations of dynamic context are also becoming more fine-grained. Prior work has incorporated multimodal signals such as gaze and hand movements, together with their temporal changes, into dynamic models that use large language models to infer user intent in real time~\cite{han2025multimodal}. Assistance systems for real-world tasks have also begun to determine how to generate responses according to the user's current environment and task state~\cite{dang2025authoring}. Location, the visual scene, user state, task progress, and interaction history can thus jointly participate in real-time generation.

These systems primarily address how to sense the current environment, understand what the user is doing, and generate a response suited to the current task or situation. However, the multiple types of information available within a single generation do not enter the final content in the same way. Even when the historical materials, live scene, and user state remain unchanged, a system can produce different content organizations. Some information may be retained in full, some may be mentioned only briefly, and some may not appear in the output at all. Existing situated generative interfaces typically make these trade-offs through a combination of post-retrieval processing, prompt constraints, and model generation, but rarely describe the resulting composition of sources in the final content. For the situated narratives studied here, this post-generation source composition is a content property that must be measured separately.

\subsection{Information Use and Source Composition in Generated Content}
\label{sec:related:grounding}

Placing information in a model's context does not mean that all of it will enter the final output to the same extent. Research on long contexts shows that even when relevant content is provided to a model, its position and organization within the context affect whether the model can use it effectively~\cite{liu2024lost}. Source documents in summarization also undergo selection, compression, and reorganization, so there is no simple one-to-one correspondence between the input materials and the final text~\cite{ravaut2024context}. Related work has also begun to analyze and intervene in a model's tendency to use parametric knowledge versus externally retrieved knowledge~\cite{wang2025knowledgeutilization}. Together, these studies show that whether information is retrieved or placed in context and how it ultimately enters the generated output are distinct questions.

Another line of research examines the correspondence between generated content and its sources. Grounding and attribution methods can determine whether a generated claim is supported by given materials and identify the specific sources that provide that support~\cite{rashkin2023measuring,gao2023citations,liu2023verifiability}. Factual content without source support is generally treated as a problem to be controlled~\cite{ji2023survey}, and efficient methods have been developed for fact-checking claims against given documents~\cite{tang2024minicheck}. In recent years, these analyses have moved to the sentence and proposition levels. Existing methods can distinguish among fully supported, partially supported, and unsupported relationships~\cite{zhang2024finegrainedcitation}; handle claims jointly supported by multiple sources~\cite{patel2024multisource}; and directly record fine-grained source relationships during generation~\cite{wei2026genprove}. It is therefore increasingly possible to describe both whether generated content is source-supported and which materials correspond to a particular claim.

Recent work has also begun to directly investigate how different knowledge sources affect generated outputs. InfoGain-RAG uses document information gain to measure the utility of an individual retrieved document to generation~\cite{wang2025infogain}. Research on parametric knowledge and external context can now analyze, and even actively alter, the degree to which models rely on different knowledge channels~\cite{wang2025knowledgeutilization,bi2026parameters}. Other studies compare models' source preferences among internal knowledge, user statements, and document statements~\cite{li2026sourcebalancing}, or analyze the contribution of different computational sources to the output through the token-generation process~\cite{lu2026tpa}. This work further establishes knowledge use, source reliance, and source contribution as distinct objects of analysis in generation.

For situated narratives, we further focus on the source composition of the final text itself. A model's strong reliance on a source does not necessarily mean that most of the final narrative comes from that source. Likewise, establishing that a claim is supported by a historical archive does not reveal how much historical information appears across the narrative as a whole. The same historical materials and situated context can produce a narrative that primarily describes the current environment or one that retains more historical information, while both outputs may remain well supported by their sources. The relevant question here is therefore not whether a source contributes, but how much content each source ultimately contributes to the complete narrative.

Source composition also cannot be inferred solely from controls specified before generation. Generative writing and creative systems provide many ways to adjust an output: users can modify requirements, change the granularity of control, and repeatedly regenerate content~\cite{mirowski2023cowriting,wu2022ai,narrativescaffolding2026,dhillon2024shaping}. However, prompt requirements do not map reliably onto specific properties of the final output~\cite{zamfirescupereira2023herding,subramonyam2025prototyping}, and model selection and the generation process itself can cause identical or similar prompts to produce different outputs~\cite{haase2026prompt}. Thus, when source composition is treated as a generation control, a preset proportion can represent only a target. The extent to which that target is realized must still be measured from the generated text.

Source information is also moving from back-end evaluation into interface design. Visualizations in human--AI collaborative writing have been used to distinguish content contributed by humans from content contributed by models~\cite{hoque2024hallmark}, while interfaces for LLM-based question answering have begun to explicitly present relationships between claims and supporting sources~\cite{martinboyle2026papertrail}. These studies show that source-level descriptions of generated outputs can support not only model diagnosis, but also interface design and user understanding.

In audio narratives for walking, source composition is further constrained by mobility itself. Content must be delivered continuously as the walk unfolds, while the user's attention is divided between the path and the surrounding environment~\cite{vazquezalvarez2011auditory}. Unlike in writing settings, users generally do not repeatedly inspect, revise, and regenerate each segment after it is produced. The system must select and organize historical materials and situated information before delivery. The text actually delivered to the user therefore reflects the realized outcome of this allocation.

This paper focuses on the realized composition of corpus knowledge, situated context, and other information in generated narratives, rather than redefining retrieval quality, factual consistency, or source attribution. We refer to the allocation of historical content specified during generation as the \emph{target grounding level}, and measure its realization in the generated output as the \emph{achieved grounding level}. The Historical Grounding Ratio (HGR) is the realized proportion of archive-supported information units (A) among all A/S/I narrative information units. Through this operationalization, we examine whether different uses of retrieved content during generation, and different allocations between historical evidence and situated information, correspond to different in-situ walking experiences when the retrieval mechanism, input types, and overall information capacity remain comparable. Section~\ref{sec:grounding} further defines HGR and describes its calculation and control during generation.
\section{Research Questions}
\label{sec:rq}

Research on location and context awareness has shown that systems can use location, user state, and the surrounding environment to determine what content to provide and when~\cite{bederson1995audio,abowd1997cyberguide,cheverst2000guide,stenton2007mediascapes,kleftodimos2023locationar}. Generative narrative systems further combine these contextual signals with external knowledge to compose new outputs at runtime~\cite{ho2025visitor,dang2025authoring,cai2025aiget,han2025multimodal,liu2025gensors}. Meanwhile, research on retrieval-augmented generation primarily uses retrieval relevance to determine whether candidate materials are suitable for use~\cite{zhao2024dense,es2024ragas,saadfalcon2024ares}, and generation attribution to assess whether final claims are supported by their sources~\cite{rashkin2023measuring,gao2023citations,liu2023verifiability,ji2023survey,zhang2024finegrained,xia2025ground}. Together, these lines of work have advanced contextual input, evidence selection, and factual reliability. They have paid less attention, however, to a question situated between retrieval and user experience: when multiple sources are relevant to the current context and can support generation, in what proportions should they enter the content ultimately delivered to users?

We focus on this output-side evidence-allocation problem by examining the realized share of historical archives in situated generative narratives. Mobile users have limited interaction time and attention, making it impossible for a system to present all retrieved materials in full~\cite{vazquezalvarez2011auditory,lanir2013the}. The proportion of historical evidence cannot be inferred directly from the number of retrieved records or represented by a target specified in a prompt, because a generative model may select, compress, and transform the input materials in different ways. Studying the experiential consequences of evidence allocation therefore requires first measuring the historical grounding actually realized in the delivered content, and then controlling and comparing it under a common retrieval procedure and comparable input capacities. We accordingly ask the following research questions:

\begin{itemize}
  \item \textbf{RQ1:} How can the share of historical evidence in a delivered narrative be operationalized as a measurable, verifiable, and controllable design variable, while remaining distinct from retrieval relevance and generation attribution?
  
  \item \textbf{RQ2:} When candidate evidence and other generation conditions are held constant, how do different levels of historical grounding affect users' experiences during real-world urban walks?
\end{itemize}

\section{Historical Grounding Ratio}
\label{sec:grounding}

The number of records returned by retrieval, a target value specified in a prompt, and whether individual claims are source-supported do not reveal the actual share of historical evidence in the information units of a delivered narrative. To make post-retrieval information allocation a measurable and comparable design variable, we introduce the Historical Grounding Ratio and describe its calculation, experimental conditions, and measurement consistency.

\subsection{Definition and Calculation of HGR}
\label{sec:grounding:hgr}

We use the Historical Grounding Ratio (HGR) to describe the share of information units in a delivered narrative that are supported by the historical corpus. HGR is not calculated from the number of historical records returned during retrieval, nor does it directly adopt the target value specified in the generation prompt. Instead, it is measured after generation from the text that pedestrians actually receive. This distinction separates the target grounding level specified before generation from the content composition realized afterward.

\begin{figure*}[!t]
  \centering
  \IfFileExists{fig_hgr_definition_annotation_v2.png}{%
    \includegraphics[
      width=0.96\textwidth,
      keepaspectratio
    ]{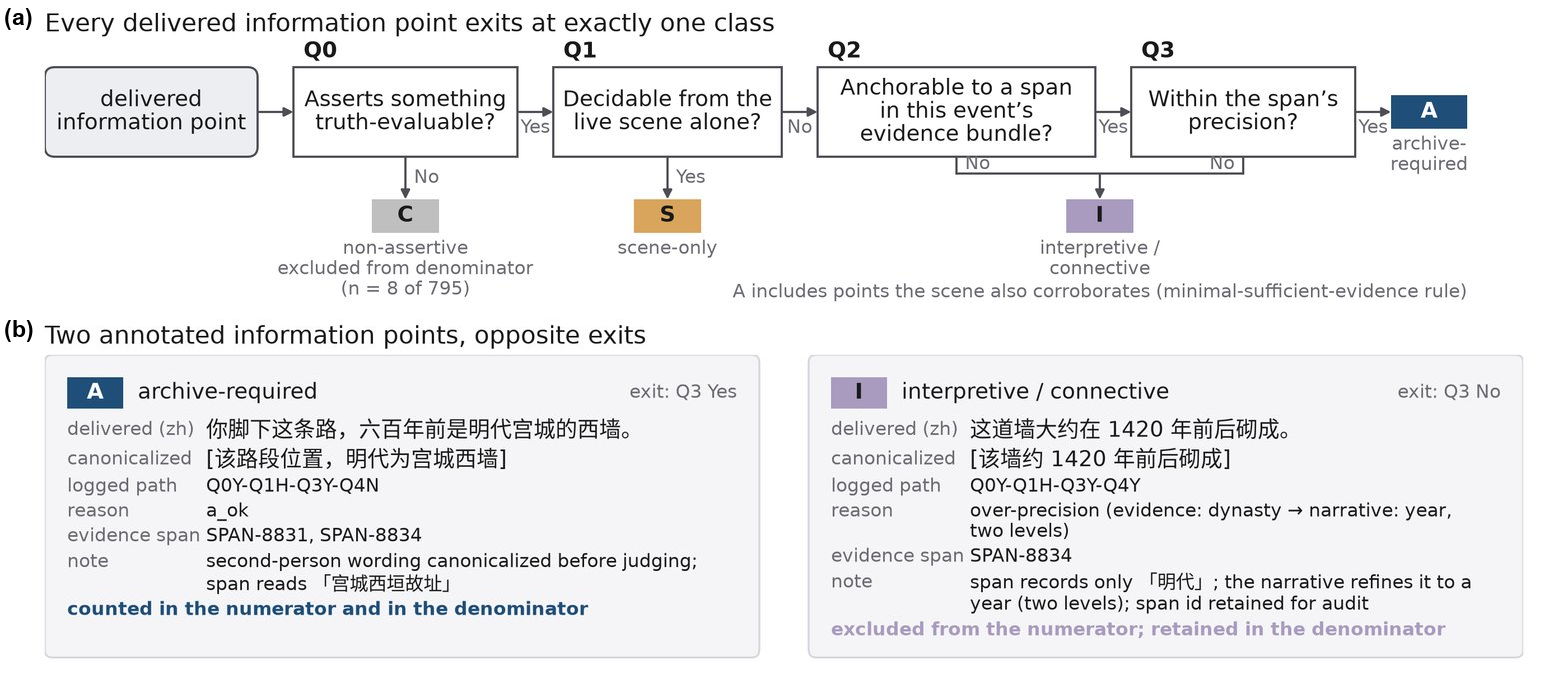}%
  }{%
    \fbox{\parbox[c][0.24\textheight][c]{0.96\textwidth}{%
      \centering Missing figure file: fig\_hgr\_definition\_annotation\_v2.png
    }}%
  }

  \caption{Classification and annotation examples for HGR information units.
  \textbf{(a)} A delivered narrative is segmented into information units whose sources can be assessed independently and classified as archive-supported, supported by the live scene, or interpretive. Category C content that does not form an independent claim is excluded from the HGR denominator.
  \textbf{(b)} Annotation examples for Category A and Category I information units, illustrating how historically supported and interpretive content are treated differently in the HGR calculation.}

  \Description{The figure shows the decision process for classifying narrative information points for Historical Grounding Ratio measurement and two annotation examples. Archive-supported information is included in the HGR numerator, while interpretive content remains in the denominator but is excluded from the numerator.}

  \label{fig:hgr-definition}
\end{figure*}

As shown in Figure~\ref{fig:hgr-definition}, we first segment each final narrative into the smallest information units whose sources can be independently assessed. Content that does not form an independent claim and serves only as an address term, grammatical connector, or transition is labeled C and excluded from the HGR calculation. The remaining narrative information units are divided into three categories according to their primary source of support:

\begin{itemize}

  \item \textbf{A---Archive-supported historical content:}
  Historical information supported by archival documents, local gazetteers, historical records, or other traceable historical sources in the retrieved evidence. If a historical information unit can also be corroborated by the current scene, it is still counted as historically grounded whenever its historical claim requires archival evidence to be established.

  \item \textbf{S---Scene-grounded situated content:}
  Situated information directly supported by the user's current location, the live visual scene, or environmental conditions, such as visible architectural elements, road configurations, materials, vegetation, or spatial relationships.

  \item \textbf{I---Interpretive/connective content:}
  Content that is not counted directly as an archival or in-situ fact, but primarily provides interpretation, association, extension of meaning, or narrative connection.

\end{itemize}

When a sentence contains multiple pieces of information whose sources can be assessed independently, we segment it by information unit rather than labeling the sentence as a whole. For example, a sentence may first describe the historical condition of a place and then identify a spatial feature that remains visible today. The former is labeled as an A unit and the latter as an S unit. Figure~\ref{fig:hgr-definition}(b) provides annotation examples for A and I units. To present source composition in greater detail, subsequent figures distinguish within Category A between A units supported only by archives and AS units also corroborated by the current scene. Both are counted in $n_A$ when calculating HGR; this subdivision does not alter the A/S/I classification or the HGR calculation.

Let $n_A$, $n_S$, and $n_I$ denote the numbers of information units in the three categories included in the calculation. For a delivered narrative text $T$, HGR is defined as:

\begin{equation}
\mathrm{HGR}(T)
=
\frac{n_A}
{n_A+n_S+n_I}.
\label{eq:hgr}
\end{equation}

HGR therefore represents the proportion of archive-supported historical content among all valid narrative information units. A higher HGR indicates that more historically supported information is retained within the limited narrative space, whereas a lower HGR indicates that more of that space is devoted to the current scene or interpretive organization. HGR describes the source composition of a narrative rather than providing an overall assessment of narrative quality; a higher value is therefore not assumed to be inherently better.

\subsection{Historical Grounding Conditions}
\label{sec:grounding:conditions}

Treating HGR as a manipulable content variable in situated storytelling depends on several information conditions.

First, the current location must be associated with sufficient historical material. When a location returns very little relevant information, the historical content of the final narrative is constrained primarily by material availability, leaving the system with few meaningful content choices. Our setting differs: location-based retrieval returns multiple historical records relevant to the current area. The generation process therefore determines not whether historical content is available, but which records and how much of their content enter the limited narrative output. Here, \emph{sufficient retrieval} does not require different locations to return identical records; it requires a candidate pool large enough to create a meaningful space for information selection and allocation.

Second, the generation process must distinguish between retrieved historical information and real-time situated information. The former comes from historical databases or knowledge corpora associated with the current location and provides knowledge about the place's past. The latter comes from camera images, the current location, and other in-situ context, and connects the narrative to the pedestrian's immediate environment~\cite{cheverst2000guide,kleftodimos2023locationar}. Although the specific content of both sources changes dynamically with location and on-site conditions, they play different informational roles during generation. Historical grounding therefore describes the content composition that these sources form after generation, rather than the quantity retrieved from either source.

Third, source allocation must occur within a relatively limited output capacity. Pedestrians must attend to both the path and the surrounding environment, which limits the time and attention available for receiving a continuous narrative~\cite{vazquezalvarez2011auditory,lanir2013the}. We constrained the number of sentence segments, target length, and information-unit quota of each narrative so that the overall information capacity remained comparable across conditions. Under this constraint, increasing the use of historical information primarily changes source composition within a finite narrative space rather than increasing historical content by indefinitely extending the text.

Based on these conditions, we established three historical grounding conditions---situation-dominant, balanced, and evidence-dominant---corresponding to low, medium, and high target levels of historical grounding. Because the candidate historical materials and live scene vary dynamically with location, we did not require every generation event to use an identical evidence bundle. The system used the same retrieval mechanism across conditions, kept the amount of candidate evidence entering generation within a comparable range, and constrained the information capacity of the final narratives. The primary difference among conditions was the extent to which historical materials and situated information were selected and organized into the final narrative after retrieval.

The situation-dominant condition organized the narrative primarily around the current scene and selected historical information from retrieved materials to supplement the place context. The balanced condition used historical materials and the situated context jointly to organize content. The evidence-dominant condition prioritized information supported by historical records, using the current scene mainly for spatial orientation and content connection. All three conditions used the same narrator identity, register, narrative scaffold, interaction method, and retrieval mechanism. They were therefore not three independent story modes, but different settings for the target use of historical materials within a narrative.

Importantly, the three conditions specify a \emph{target grounding level} during generation, whereas HGR is measured from the text actually delivered after generation. The generative model, the specific retrieved content, and the real-time context may all cause the final output to deviate from its target setting. We therefore avoid fixed-ratio labels such as H30, H60, or H90. Instead, we recalculate the achieved HGR of every narrative after generation and test in Section~\ref{sec:results:manip} whether the three conditions produced the expected realized differences.

\subsection{HGR Measurement and Consistency}
\label{sec:grounding:annotation}

Each narrative delivered during the study was exported together with the evidence bundle and scene description available when it was generated. Source classification was therefore based on these two inputs rather than on an annotator's own world knowledge. We developed an annotation protocol specifying how to segment information units and distinguish among the A/S/I provenance categories. A machine annotator applied this protocol to all 54 final narrative segments, producing 793 information units. Eight Category C units that did not form independent claims were excluded from the HGR denominator, leaving 785 valid A/S/I units. HGR for each segment was calculated from the counts in Equation~\ref{eq:hgr}. All 54 segments retained their complete nine-sentence narrative texts and had complete ratings and source annotations.

Two coders independently annotated a 20.4\% subsample of the segments using the same A/S/I classification scheme as the machine annotator. Their segment-level HGR values showed high agreement (Pearson $r = .984$; $\mathrm{ICC}(A,1) = .965$, two-way random effects, absolute agreement, single measurement; mean absolute difference $= .038$). Using the mean of the two coders as the segment-level consensus value, the machine annotations were strongly correlated with the human values ($r = .940$) and showed a small negative bias ($M = -.037$, $\mathrm{MAE} = .053$), indicating that the machine slightly underestimated HGR. Human consensus values in the subsample ranged from $.303$ to $.892$, while values across all 54 segments ranged from $.143$ to $.895$. The human-verified sample therefore covered the main range of the measure.

All statistical analyses used HGR values produced by the machine annotation of the 54 segments. The double-coded subsample was used to assess how closely these annotations matched human coding, rather than to replace a subset of the machine-generated values. This separation is necessary because human coding covered only one fifth of the material. Replacing that subset with human values while retaining machine values for the remainder would construct a single variable from two annotation sources, allowing between-condition differences to reflect the distribution of annotation sources rather than the narratives themselves. Retaining a single annotation source and separately reporting its deviation from human judgment provides a more testable procedure. As shown in Section~\ref{sec:results:manip}, the direction of this bias weakened rather than strengthened the experimental manipulation.

Two limitations should be stated explicitly. First, the intraclass correlation coefficient reflects agreement in continuous segment-level HGR values, not agreement in the assignment of A/S/I category labels. The three annotations used different segmentation granularities and produced 215, 197, and 171 information units, respectively. Calculating category-level $\kappa$ or $\alpha$ would require information-unit labels aligned across coders, which the existing records do not provide. Second, the subsample was matched to the corresponding segments after analysis rather than selected in advance through stratified sampling by experimental condition. It should therefore be treated as a general subsample.

This definition and measurement procedure enables historical grounding to be quantified in actual generated texts. The next section describes how GeoDrama organizes location-based retrieval, real-time scene input, and narrative generation in a dynamic walking context, and how it implements the historical grounding conditions described above.

\section{GeoDrama System}
\label{sec:system}

The preceding section makes it possible to measure HGR from final narratives. Comparing different HGR configurations during real-world walks, however, also requires an operational platform that can control information allocation and record the complete generation process. We therefore developed GeoDrama, which treats post-retrieval information allocation as a manipulable generation strategy within a unified pipeline for location sensing, scene understanding, and historical retrieval.

\begin{figure*}[!t]
  \centering
  \includegraphics[width=\textwidth]{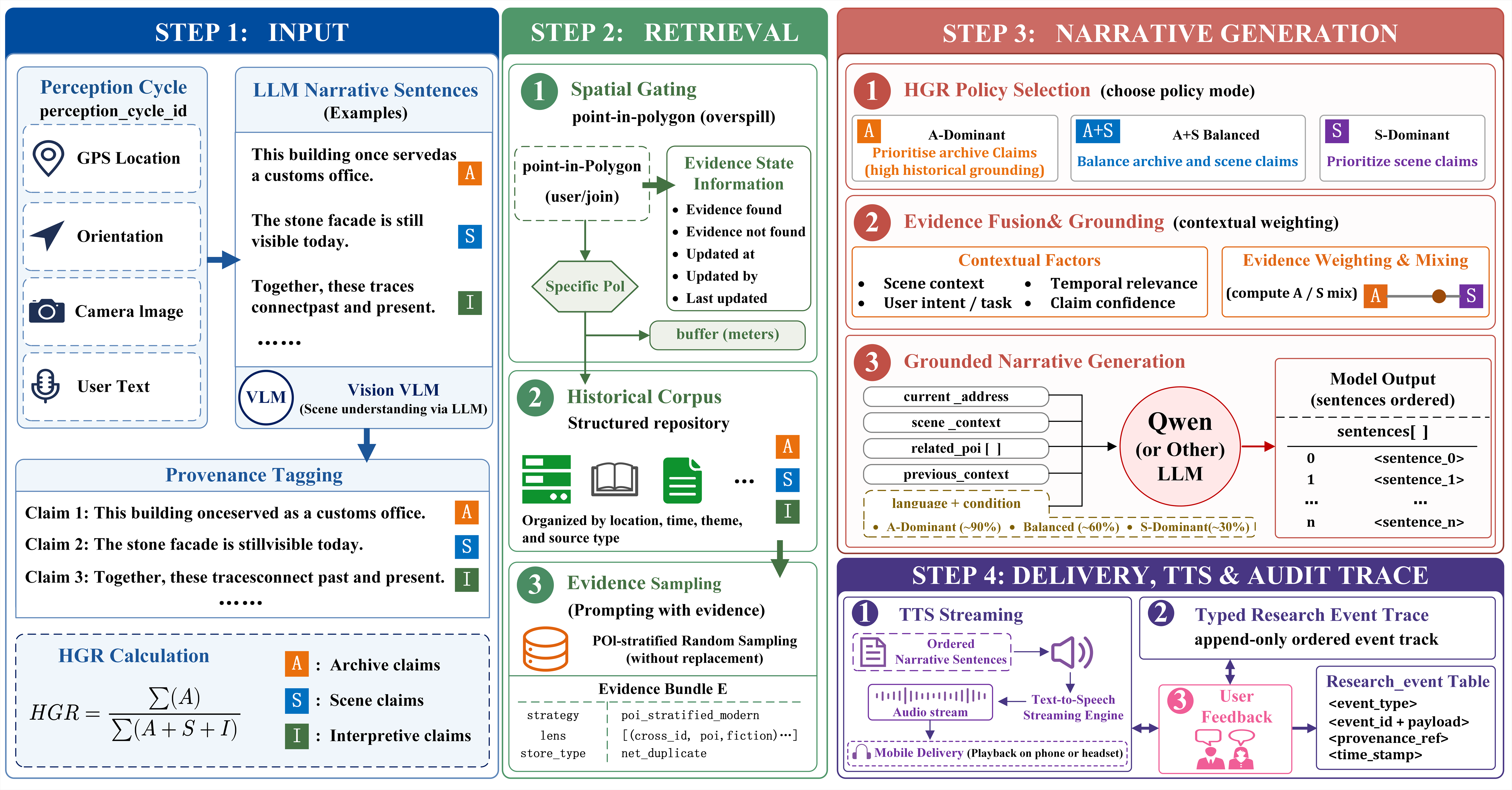}

  \caption{GeoDrama's end-to-end architecture and data flow.
  \textbf{Step 1---Input:}
  GPS location, orientation, camera imagery, and user text support scene understanding, provenance tagging, and HGR calculation.
  \textbf{Step 2---Retrieval:}
  Spatial gating identifies the active place context, after which the historical corpus is queried and evidence is sampled into a traceable bundle.
  \textbf{Step 3---Narrative Generation:}
  A target allocation policy guides evidence fusion and grounding, and the language model produces an ordered sequence of narrative sentences.
  \textbf{Step 4---Delivery, TTS, and Audit Trace:}
  The ordered sentences are streamed as audio for mobile delivery, while typed research events and post-experience feedback are recorded for auditing.}

  \Description{GeoDrama system architecture organized into four connected steps. Step 1 combines GPS location, orientation, camera imagery, and user text with scene understanding, provenance tagging, and HGR calculation. Step 2 applies spatial gating, accesses the historical corpus, and samples a traceable evidence bundle. Step 3 applies a target allocation policy, fuses and grounds evidence, and generates ordered narrative sentences. Step 4 streams the sentences through text-to-speech for mobile delivery and records typed research events and post-experience feedback.}

  \label{fig:pipeline}
\end{figure*}

GeoDrama is a mobile research prototype designed for real-world walking. As a user walks, GeoDrama senses the current location and surrounding streetscape, retrieves historical archival records associated with that location, generates a situated historical narrative, and presents it through sentence-level speech and location-linked visual content. Figure~\ref{fig:pipeline} shows this end-to-end loop in four connected steps: input and provenance tagging, spatially constrained retrieval and evidence sampling, grounded narrative generation under a target allocation policy, and mobile delivery with TTS and an auditable research-event trace.

Location and live-scene input serve different functions at different stages of the pipeline. Location determines which historical records are geographically eligible for retrieval. The visual description of the streetscape is used both to semantically rerank those records and as contextual input to narrative generation. The visible scene therefore participates first in evidence selection and then in narrative organization.

Between these two stages, GeoDrama exposes information allocation as a separate, controllable strategy during generation. Candidate evidence changes dynamically with location and the in-situ context, but all three experimental conditions use the same sensing and retrieval mechanisms. The conditions primarily vary the relative extent to which archival, situated, and interpretive content enters the final narrative after retrieval. GeoDrama's processing chain thus comprises evidence selection, information allocation, and narrative generation, with information allocation manipulated through the HGR conditions.

\subsection{Location and Scene}
\label{sec:system:perception}

GeoDrama continuously samples GPS coordinates and performs a point-in-polygon test between each location and a manually constructed layer of historical spatial polygons in the study area. It retains every polygon containing the current location rather than selecting only the nearest one. This approach accommodates the spatial overlap common in historical cities: a single location may simultaneously belong to an extant building, the former extent of an institution, a lane that has since disappeared, and a larger named place. Spatial containment therefore determines the set of historical records eligible for retrieval at the current location. Figure~\ref{fig:corpus} shows the composition of the historical spatial polygons in the study area and the set of polygons activated by a single location.

GeoDrama uses historical spatial polygons rather than simple distance radii because the extents of historical places are rarely circular and do not necessarily coincide with present-day address boundaries. The mobile interface displays a reverse-geocoded address to help pedestrians orient themselves, but this address is not used as an association key or selection criterion for historical records. Spatial association relies only on the current coordinates and stable polygon identifiers.

In parallel with location sampling, GeoDrama continuously captures camera images and uses a vision--language model to convert the most recent available frame into a short, constrained textual scene description. This description includes observable elements required for subsequent retrieval and generation, such as architectural components, road configurations, vegetation, material surfaces, visible activities, and spatial relationships. Both retrieval and narrative generation use this textual representation; raw camera images are not passed directly to the narrative generator. Scene-supported information units can therefore be verified against the scene description retained at generation time without revisiting participants' raw camera images.

\subsection{Retrieval and Scene-Based Reranking}
\label{sec:system:retrieval}

Evidence retrieval proceeds in two stages. The first stage applies a geographic constraint: every historical polygon containing the pedestrian's current location returns its associated archival records, forming a geographically eligible candidate pool with up to 20 records retained for each point of interest. Across the 54 delivered segments, the candidate pool for each generation contained 110--358 records, with a median of 183.

The second stage performs scene-based semantic reranking. A lightweight Chinese sentence-embedding model (\texttt{BAAI/bge-small-zh-v1.5}) encodes the current scene description, which is then compared by cosine similarity with precomputed embeddings of the candidate records. The candidates are reranked within the geographically constrained pool, and the ten highest-ranked records form the evidence bundle $E$ passed to the generator. This bundle contains 776--886 Chinese characters of archival text ($M \approx 830$).

This ordering is consequential. Reranking occurs strictly within the candidate pool selected by the geofence. Visual relevance can therefore change only the priority of materials already permitted by the current location. If the polygon associated with a record does not contain the pedestrian, the record cannot be promoted regardless of its semantic similarity to the current streetscape. Relevance can reorder the history that the location permits the system to consider, but it cannot expand that geographic boundary.

Each retrieval event logs the location and polygon identifiers, candidate records, scene description used as the query, similarity scores, final evidence bundle, and record order before and after reranking. The mechanism can therefore be directly audited rather than merely inferred from the system architecture. Scene-based semantic reranking succeeded in 43 of the 54 segments, with a mean latency of 9.44\,ms. In every one of these 43 segments, the order of the ten-record evidence bundle changed relative to its pre-reranking order. No scene description was available at runtime for the remaining 11 segments, so retrieval fell back to deterministic random ordering within the geofenced candidate pool. These 11 segments comprised five situation-dominant, five evidence-dominant, and one balanced segment, and most occurred within the first two segments of a walk. The fallback therefore did not systematically coincide with the experimental manipulation.

\subsection{Generation and Output}
\label{sec:system:generation}

The narrative generator receives four inputs: the evidence bundle $E$ retrieved for the current location, the real-time scene description, a fixed nine-sentence narrative scaffold, and an evidence-allocation instruction corresponding to the experimental condition. Narratives are generated by a locally deployed, Chinese-capable, 8-billion-parameter open-weight model (\texttt{qwen/qwen3-8b}). All three conditions use the same model and identical decoding parameters, narrator identity, and target number of sentences. The instruction constrains the number of information units in a segment rather than its character length. Across the 54 final segments, mean generation latency was 4519\,ms, with a range of 3309--11366\,ms.

In addition to text and sentence-level speech output, GeoDrama uses the current scene information and generated narrative to produce visual content. It maintains the spatial association between each generated visual segment and its trigger location, forming a spatially augmented narrative sequence organized along the walking route.

The experimental condition enters the generation process only through this instruction, which specifies how a segment should be organized and how many of its information units should carry archival content. All retrieved records remain available in all three conditions; no records are filtered before generation according to condition. The system logs both the number of information units requested by the instruction and the records the model reports having used. The former is only a system input and the latter only a model self-report; neither is treated as a measurement of the final text. Achieved HGR is obtained only through the annotation procedure described in Section~\ref{sec:grounding:annotation}.

Output uses sentence-level streaming speech synthesis, allowing pedestrians to hear the first sentence while the remaining sentences are still being generated and to keep their attention on the street. The study logger records every sensing cycle, retrieval, generation, and questionnaire response as a typed event under a common collection-batch identifier. All data reported in this paper were selected using this identifier.

\section{Field Study}
\label{sec:study}

To examine whether different HGR configurations produce perceptible experiential differences during real-world walking, we deployed GeoDrama in an archive-rich historical district in Nanjing and conducted a within-subject field study. Eighteen participants walked along a predefined route and experienced situated historical narratives generated under the situation-dominant, balanced, and evidence-dominant HGR conditions. Condition order and route segment were balanced through complete counterbalancing. We combined post-segment ratings, retrospective comparisons after the walk, and semi-structured interviews to examine how different information configurations affected place relevance, historical understanding, scene integration, information fit, and intention to explore further. The following sections describe the participants and study area, experimental conditions and tasks, procedure, measures, and analysis plan.

\subsection{Participants and Study Area}
\label{sec:study:participants_setting}

We recruited participants through on-site intercepts in the study area. Researchers invited eligible pedestrians to participate voluntarily and sought variation in age and gender while including both local residents and visitors. Eligibility criteria were being at least 18 years old, volunteering to participate, being able to complete an outdoor walk independently, and being able to operate the mobile interface without continuous assistance from a researcher. Data were included only when a participant completed all three experimental conditions and their corresponding questionnaires and successfully received narrative content in all three conditions.

Eighteen eligible participants took part. All completed the full study without withdrawing, and data from all 18 were included in the analysis. Participants included 11 women and 7 men, aged 18--50. Because some reported an age range or approximate age, we do not calculate an exact mean or standard deviation for age. The walking experience in each condition lasted approximately 8 minutes, for approximately 24 minutes across all three conditions. The post-study semi-structured interview lasted approximately 10 minutes, in addition to the system demonstration and post-segment questionnaires.

The study took place in a historical district in central Nanjing. The historical corpus used by the system was curated for this geographic area and spatially aligned with corresponding historical polygons. The area therefore served both as the physical setting for the walking task and as the spatial extent from which the system retrieved local historical information. All sessions were conducted under weather conditions suitable for outdoor walking.

\begin{figure}[!t]
  \centering
  \IfFileExists{corpus_geofence.png}{%
    \includegraphics[width=\columnwidth]{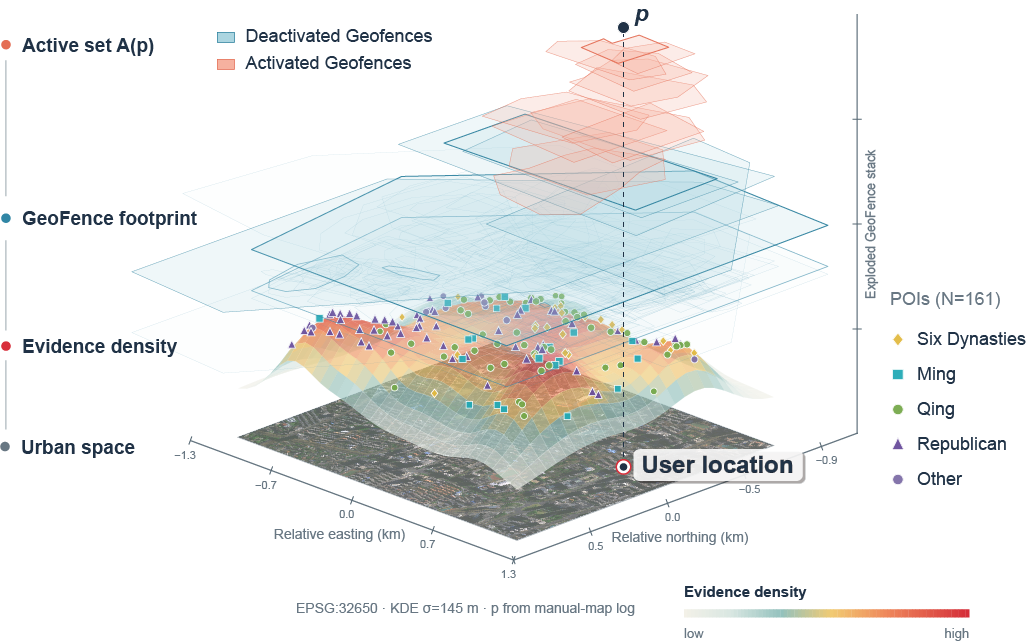}%
  }{%
    \fbox{\parbox[c][0.35\textheight][c]{\columnwidth}{%
      \centering Missing figure file: corpus\_geofence.png
    }}%
  }
  \caption{Spatial organization of the archival corpus in the study area and the active spatial context formed by a single location.
  \textbf{Bottom layer:} Kernel-density distribution of 161 points of interest aligned with the study area (Gaussian kernel, $\sigma=145$\,m; EPSG:32650), with point markers distinguished by historical period.
  \textbf{Middle layer:} The 161 true-scale historical spatial polygons associated with these points of interest. The polygons overlap and nest rather than forming a disjoint spatial partition.
  \textbf{Top layer:} A point-in-polygon test for an observed location $p$ on the study route. $\mathrm{covers}(p,\mathrm{GeoFence})$ returns all 18 polygons containing $p$, forming the active spatial context $A(p)$; six represent larger contextual places and twelve represent specific physical features.}
  \Description{A three-layer exploded axonometric diagram of the study area. The bottom layer shows a kernel density surface of 161 points of interest, coloured from low to high density, with point markers distinguished by historical period, including Six Dynasties and Southern Tang, Ming, Qing, Republican, and other periods. The middle layer shows 161 true-scale historical geofence polygons that overlap and nest within one another. The top layer shows a vertical dashed line rising from a GPS position p on the study route through the polygon stack, with 18 polygons highlighted as the active spatial context returned by a point-in-polygon test. Six polygons represent larger contextual places and twelve represent specific physical features.}
  \label{fig:corpus}
\end{figure}

The system-aligned corpus subset for the study area contained 161 points of interest and 161 associated true-scale historical spatial polygons, whose spatial distribution substantially overlapped the experimental route. These polygons do not partition the area into mutually exclusive units. Instead, they overlap and nest, allowing a single location to correspond simultaneously to historical places at multiple spatial scales (Figure~\ref{fig:corpus}).

For the observed location shown in Figure~\ref{fig:corpus}, the point-in-polygon test returned 18 polygons containing that position: six larger contextual places and twelve specific physical features. During operation along the experimental route, the resulting candidate pools contained 110--358 records, with a median of 183. These values define the spatial and corpus environment of the field study.

\subsection{Experimental Conditions and Task}
\label{sec:study:conditions}

We used a one-factor, three-level within-subject design, with HGR condition as the only manipulated factor. In the situation-dominant condition, a smaller share of the narrative was allocated to archive-supported content and more use was made of real-time scene information at the participant's location. The balanced condition used a relatively even information configuration between archival evidence and the in-situ context. The evidence-dominant condition allocated a larger share of the generated narrative to archive-supported content.

Each participant experienced all three conditions once during a continuous walk. Their task was to follow the predefined route at their own pace and receive situated historical narratives related to their current location through the mobile interface. The route was divided into three consecutive segments of approximately equal walking duration. Each segment contained multiple geofence polygons, and one HGR condition controlled the information configuration of all narratives within that segment. Route segments were determined before data collection based on spatial extent and the geofence layer, without reference to narratives generated during the study.

All six possible presentation orders of the three conditions were used, with three participants assigned to each order. Each condition therefore appeared six times in each route segment and six times in each presentation position, preventing any systematic correspondence between a condition and a particular route segment or order. Although the three segments differed in their specific places and historical content, complete counterbalancing balanced these differences across participants.

We chose a within-subject design for two reasons. First, participants may differ substantially in their prior preferences for the amount of historical information, narrative style, and walking experience. Having each participant experience every condition reduces the influence of these individual differences on condition comparisons. Second, because we examine how participants judge the suitability of different information configurations for walking, first-hand experience of all three conditions provides a common basis for the final retrospective comparisons.

\subsection{Procedure}
\label{sec:study:procedure}

Before the formal study began, participants read the study information and provided electronic informed consent. Researchers informed them that they would receive generated historical narratives related to places along the route, organized from archival records associated with those places and in-situ contextual information, and that the narratives might contain errors. Participants were not told that the three route segments corresponded to different HGR conditions or informed of our expectations regarding the information configurations.

\begin{figure}[!t]
  \centering
  \IfFileExists{interface.png}{%
    \includegraphics[width=\columnwidth]{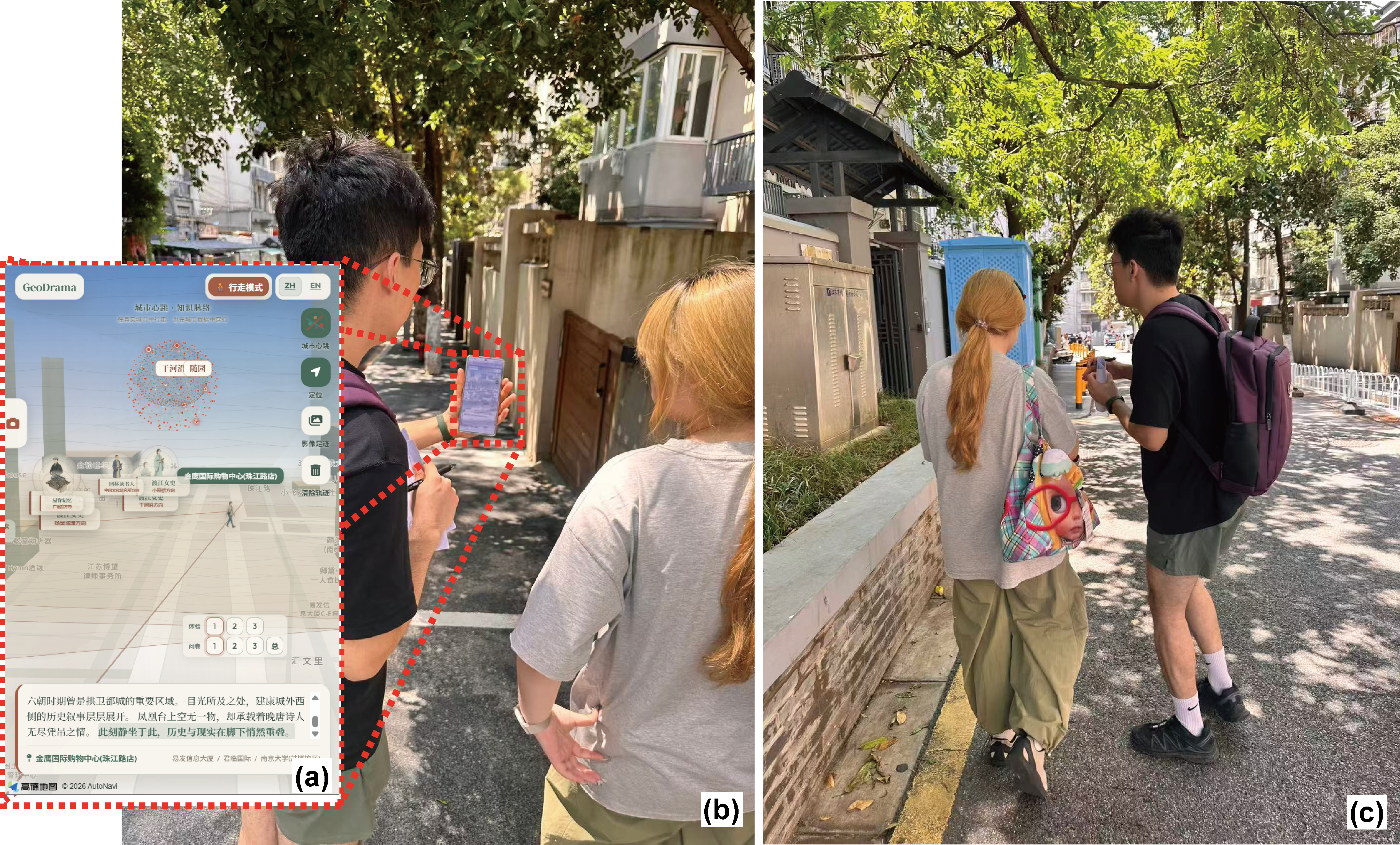}%
  }{%
    \fbox{\parbox[c][0.30\textheight][c]{\columnwidth}{%
      \centering Missing figure file: interface.png
    }}%
  }
  \caption{GeoDrama's mobile interface and field-study procedure.
  \textbf{(A)} The mobile interface used in the study, displaying the current location, historical spatial information, and generated narrative content.
  \textbf{(B)} Before the formal walk, a researcher demonstrates system operation and content delivery to a participant.
  \textbf{(C)} During the formal experience, a participant walks along the experimental route and receives situated historical narratives through the mobile interface.}
  \Description{Three-panel field-study figure. Panel A shows the GeoDrama mobile interface used during the study, including the participant's current location, historical spatial information, and generated narrative content. Panel B shows a researcher demonstrating the interface before the formal walk. Panel C shows a participant walking along the study route and receiving situated historical narratives through the mobile interface.}
  \label{fig:interface}
\end{figure}

After consent, a researcher used the study's mobile interface to demonstrate location sensing, narrative triggering, and content delivery (Figure~\ref{fig:interface}B). The demonstration covered interface operation only and did not explain the experimental conditions or narrative content. Figure~\ref{fig:interface}A shows the mobile interface used in the study, including the main areas displaying the current location, historical spatial information, and generated narrative.

After the demonstration, participants entered the formal route using their own phones and independently completed the three consecutive segments at their own pace (Figure~\ref{fig:interface}C). They could stop, inspect content, or take photographs as desired. A researcher accompanied them from several meters away, provided no supplementary explanation of the narratives, and intervened only in the event of a location, network, or interface failure, thereby reducing the influence of researcher presence on the narrative experience.

At the end of each route segment, participants stopped before entering the next segment and evaluated the narrative they had just experienced using five seven-point items. After completing all three segments, they took part in an approximately 10-minute semi-structured interview about which narratives they remembered, where the amount of information seemed excessive or insufficient, and whether any content appeared questionable. They then completed a final comparison questionnaire, selecting from the three experienced narrative conditions the one they preferred overall, the one with the most appropriate amount of historical information, and the one that most strengthened the perceived connection between the narrative and the current place.

\subsection{Measures}
\label{sec:study:measures}

We collected four types of measures: immediate experience ratings after each route segment, retrospective comparisons after all three segments, open-ended text and interview feedback, and achieved HGR calculated from the final delivered narratives.

\paragraph{Post-segment experience ratings.}
At the end of each route segment, participants evaluated the narrative they had just experienced using five single-item measures: place relevance (Q1), historical understanding (Q2), scene integration (Q3), information fit (Q4), and intention to explore further (Q5). Table~\ref{tab:measures} reports the items shown to participants.

\begin{table}[!t]
  \caption{Experience measures administered after each route segment. All items used a seven-point scale.}
  \label{tab:measures}
  \centering
  \small
  \begin{tabular}{@{}lp{0.21\columnwidth}p{0.58\columnwidth}@{}}
    \toprule
    ID & Dimension & Item shown to participants \\
    \midrule
    Q1 & Place relevance &
    How relevant was the content you just heard to the specific place where you were standing? \\

    Q2 & Historical understanding &
    After hearing this segment, how clearly did you understand what this place was like in the past? \\

    Q3 & Scene integration &
    How closely was the content you just heard connected to the physical scene in front of you? \\

    Q4 & Information fit &
    How appropriate was the amount of information in this segment for you? \\

    Q5 & Intention to explore further &
    After hearing this segment, how much did you want to continue exploring this area? \\
    \bottomrule
  \end{tabular}
\end{table}

All items were rated from 1 to 7, with verbal anchors only at 1, 4, and 7: 1 (not at all), 4 (moderately), and 7 (very much). Points 2, 3, 5, and 6 were unlabeled. Higher scores on Q1, Q2, Q3, and Q5 respectively indicated stronger perceived place relevance, clearer understanding of the place's past, closer integration between the narrative and the visible scene, and stronger intention to continue exploring the surrounding area. Q4 was scored in its original direction: a higher score indicated that the participant considered the amount of information more appropriate, not that the narrative contained more information. Single-item measures reduced interruptions caused by repeated stopping and questionnaire completion during the street walk.

\paragraph{Retrospective comparisons.}
After completing the three route segments, participants answered three forced-choice questions:

\begin{enumerate}
  \item Which of the three experiences did you prefer overall?
  \item Which narrative provided the most appropriate amount of information?
  \item Which narrative most strongly made you feel that the history was about the place in front of you?
\end{enumerate}

The response options for each question were A, B, and C, referring respectively to the first, second, and third route segments experienced by that participant. These labels were not fixed to the situation-dominant, balanced, or evidence-dominant conditions. The system mapped each selection back to the corresponding HGR condition according to the participant's assigned order. Retrospective comparisons captured relative judgments after experiencing all conditions and were not combined with the immediate post-segment ratings.

The final questionnaire also included an open-ended question: Why did you prefer this segment? Responses recorded participants' reasons for their overall preference and provided qualitative material for interpreting the retrospective choices.

\paragraph{Interview feedback.}
The semi-structured interview covered five predefined themes: participants' responses to the generated narratives, their overall walking experience, their willingness to continue using this form of narrative, changes in their perception of the space before and after the experience, and their views of the research project. Interviews used open-ended follow-up questions without quantitative ratings. Interview recordings, researchers' field notes, and open-ended responses from the final questionnaire jointly formed the qualitative feedback. These materials were used to interpret experiential differences in the quantitative results; they were not included in statistical tests or used to estimate the prevalence of particular views.

\subsection{Data Analysis}
\label{sec:study:analysis}

The 18 participants each completed three experimental conditions, producing 54 participant--condition observations and 270 single-item rating responses. After the walk, each participant completed three forced-choice questions, producing 54 retrospective-choice responses.

To answer RQ1, we first report the distribution of achieved HGR under the three allocation conditions and test whether the situation-dominant, balanced, and evidence-dominant conditions produced the intended grounding gradient. Because achieved HGR is calculated from the final delivered text, this manipulation check also indicates whether the information configurations specified by the experimental instructions were realized in the output.

To answer RQ2, the main analysis treated the fully counterbalanced experimental condition as the independent variable and separately tested its effect on the five experiential outcomes Q1--Q5. Because each outcome was measured using a single seven-point item and each participant contributed only three repeated observations per item, we conducted a Friedman test for each item and used Wilcoxon signed-rank tests for pairwise comparisons among conditions. The three pairwise comparisons within each item were adjusted using the Holm method~\cite{friedman1937ranks,wilcoxon1945ranking}. We report Kendall's $W$ for omnibus condition effects and rank-biserial $r$ for pairwise comparisons.

To examine whether place relevance changed differently from the other experiential outcomes, we conducted an exploratory within-subject contrast. For each participant, we calculated the evidence-dominant minus situation-dominant difference for Q1 and subtracted the mean of the corresponding differences for Q2--Q5, then tested the resulting contrast scores using a Wilcoxon signed-rank test. This analysis tests whether the condition-related change in Q1 differs from the overall direction of change in the other outcomes; the Q2--Q5 mean is not interpreted as an independent psychological construct or validated composite scale. Because Q2--Q5 used the same seven-point scale and scoring direction, we additionally examined their internal consistency and used their mean as an exploratory cross-dimensional experience composite. We analyzed this composite using a Friedman test and Holm-adjusted Wilcoxon pairwise comparisons. To test whether the balanced condition exceeded the mean of the two extreme conditions, we constructed the quadratic contrast $c_i=2\times\mathrm{balanced}_i-\mathrm{situation}_i-\mathrm{evidence}_i$ and applied an exact sign-flipping permutation test, with a Wilcoxon test and leave-one-out analyses as robustness checks. This composite was used only to explore shared cross-dimensional variation.

To determine whether differences between the two extreme conditions on each experiential dimension were smaller than a practically meaningful effect, we conducted paired-sample equivalence tests (TOST) comparing the evidence-dominant and situation-dominant conditions, using $d_z=0.70$ as the equivalence bound.

Qualitative materials were used to interpret the main patterns in the quantitative results. For each of the three retrospective choices, we counted selections and proportions by condition and used an exact multinomial goodness-of-fit test to compare the observed distribution against the theoretical distribution in which all three conditions were equally likely to be selected. Because these choices were made after participants completed all conditions, we interpret them as a complementary comparison to the immediate post-segment ratings. Open-ended questionnaire responses, interview recordings, and field notes were organized according to the five themes defined by the interview guide: responses to the narratives, the walking experience, willingness to use the system again, changes in spatial experience, and views of the research project.

\section{Results}
\label{sec:results}

All 18 participants completed all three conditions, yielding $18 \times 3 \times 5 = 270$ ratings with no missing values. All 54 generated narrative segments retained complete text and had complete ratings and source annotations. Source labels were produced by the annotation model; the manual verification procedure and agreement results for 11 segments are reported in the Methods section.

\begin{figure}[!t]
  \centering
  \includegraphics[
    width=\columnwidth,
    keepaspectratio
  ]{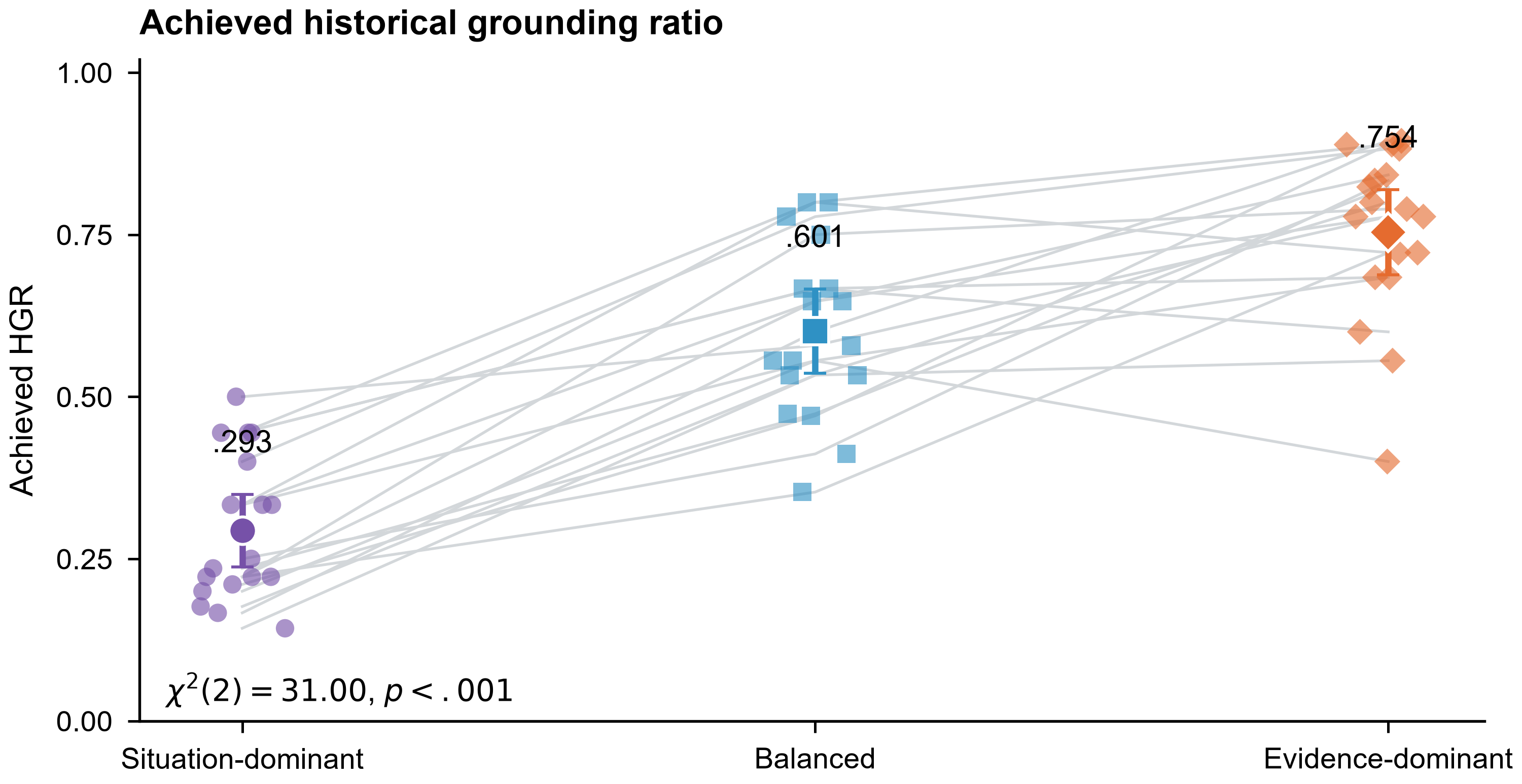}
  \caption{Achieved historical grounding ratio (HGR) under the three experimental conditions. Points show segment-level values; thick horizontal marks and vertical intervals show condition means and 95\% confidence intervals.}
  \Description{Segment-level historical grounding ratios under the situation-dominant, balanced, and evidence-dominant conditions, with means and confidence intervals.}
  \label{fig:hgr-achieved}
\end{figure}

\subsection{Manipulation Check}
\label{sec:results:manip}

The three historical grounding conditions produced distinct achieved HGR values as intended (Table~\ref{tab:manipcheck}; Figure~\ref{fig:hgr-achieved}). Mean segment-level HGR was .293 in the situation-dominant condition, .601 in the balanced condition, and .754 in the evidence-dominant condition. A Friedman test showed a significant condition effect, $\chi^2(2)=31.00$, $p<.001$. All three Wilcoxon pairwise comparisons remained significant after Holm correction, with effect sizes of $r=.88$, $.88$, and $.68$.

\begin{figure}[!t]
  \centering
  \includegraphics[
    width=\columnwidth,
    keepaspectratio
  ]{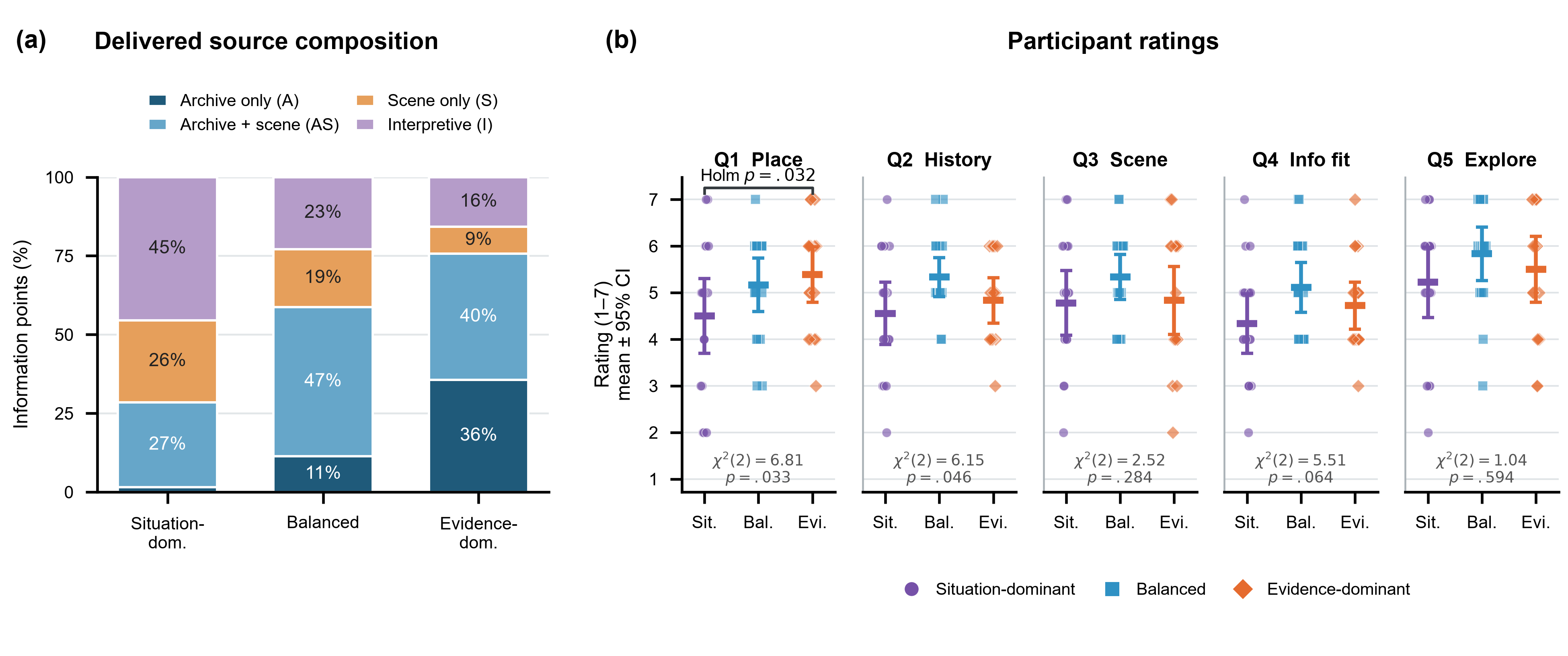}
  \caption{Source composition and participant ratings under the three historical grounding conditions.
  \textbf{(a)} Pooled proportions of archive-only (A), archive-and-scene-supported (AS), scene-only (S), and interpretive (I) information units.
  \textbf{(b)} Participant ratings for place relevance (Q1), historical understanding (Q2), scene integration (Q3), information fit (Q4), and intention to explore further (Q5). Points show individual ratings; thick horizontal marks and vertical intervals show condition means and 95\% confidence intervals.}
  \Description{Two side-by-side panels. The left panel shows stacked source-composition bars for the three conditions. The right panel shows participant-level ratings and condition means for the five in-situ rating items.}
  \label{fig:source-composition-ratings}
\end{figure}

As historical grounding increased, the combined proportion of A and AS increased while S and I decreased (Figure~\ref{fig:source-composition-ratings}(a)). The figure shows pooled source composition by condition, while Table~\ref{tab:manipcheck} reports condition means for segment-level HGR.

This shift was not achieved by increasing the total amount of narrative content. The situation-dominant, balanced, and evidence-dominant conditions contained 264, 254, and 267 valid information units, respectively, corresponding to means of 14.67, 14.11, and 14.83 units per segment. Narratives in all three conditions contained nine sentences. The manipulation therefore operated primarily by changing source composition within a comparable amount of discourse rather than simply increasing the number of information units or sentences.

\begin{table}[t]
\centering
\caption{Manipulation check. Achieved HGR is based on all 54 narrative texts.}
\label{tab:manipcheck}
\small
\begin{tabular}{lccc}
\toprule
 & Situation-dominant & Balanced & Evidence-dominant \\
\midrule
Achieved HGR $M$ (SD)
& .293 (.113) & .601 (.131) & .754 (.132) \\
Total valid information units
& 264 & 254 & 267 \\
Mean information units per segment
& 14.67 & 14.11 & 14.83 \\
\bottomrule
\end{tabular}
\end{table}

\subsection{Perceived Relevance Between Narrative Content and Place}
\label{sec:results:q1}

Among the five individual ratings, Q1 was the only item with a pairwise comparison that remained significant after within-item Holm correction. Mean ratings increased from the situation-dominant to the balanced and evidence-dominant conditions, Friedman $\chi^2(2)=6.81$, $p=.033$, Kendall's $W=.19$. Of the three pairwise comparisons, only the evidence-dominant versus situation-dominant difference remained significant after Holm correction: median difference $=1.0$, $z=2.55$, $p_{\mathrm{Holm}}=.032$, rank-biserial $r=.71$, 95\% CI $[.30,.95]$, $d_z=0.70$ (Figure~\ref{fig:source-composition-ratings}(b)).

Because the three conditions contained similar numbers of information units and sentences, this difference corresponds to a change in source composition within a comparable amount of discourse rather than an increase in total content. The Holm correction above treats the three pairwise comparisons within each item as the correction family. If all 15 comparisons across the five items are treated as a single family, the evidence-dominant versus situation-dominant difference on Q1 is no longer significant.

\subsection{Composite Results for the Remaining Experience Dimensions}
\label{sec:results:q2q5}

The individual results for Q2--Q5 produced no significant pairwise differences after correction (Table~\ref{tab:ratings}; Figure~\ref{fig:source-composition-ratings}(b)). The omnibus Friedman test for historical understanding (Q2) was significant, but none of its three pairwise comparisons passed within-item Holm correction (minimum $p_{\mathrm{Holm}}=.108$).

\begin{table*}[t]
\centering
\caption{Condition effects for the five in-situ rating items ($n=18$, within-subject; Q4 is scored in its original direction, with higher values indicating better fit). The table reports the omnibus Friedman test for each item. Pairwise comparisons use Wilcoxon signed-rank tests with within-item Holm correction.}
\label{tab:ratings}
\small
\setlength{\tabcolsep}{4pt}
\renewcommand{\arraystretch}{1.1}
\begin{tabular}{@{}llccccccp{5.4cm}@{}}
\toprule
& & \multicolumn{3}{c}{$M$ (SD)} & \multicolumn{3}{c}{Friedman} & \\
\cmidrule(lr){3-5}\cmidrule(lr){6-8}
Item & Dimension & Situation-dominant & Balanced & Evidence-dominant
& $\chi^2(2)$ & $p$ & $W$ & \parbox[t]{5.4cm}{\raggedright Significant corrected pairwise comparison} \\
\midrule
Q1 & Content--place relevance
& 4.50 (1.62) & 5.17 (1.15) & 5.39 (1.20)
& 6.81 & .033 & .19
& Evidence-dominant $>$ situation-dominant ($p_{\mathrm{Holm}}=.032$) \\

Q2 & Historical understanding
& 4.56 (1.34) & 5.33 (0.84) & 4.83 (0.99)
& 6.15 & .046 & .17
& None (minimum $p_{\mathrm{Holm}}=.108$) \\

Q3 & Scene integration
& 4.78 (1.40) & 5.33 (0.97) & 4.83 (1.47)
& 2.52 & .284 & .07
& None \\

Q4 & Information fit
& 4.33 (1.28) & 5.11 (1.08) & 4.72 (1.02)
& 5.51 & .064 & .15
& None \\

Q5 & Intention to explore further
& 5.22 (1.52) & 5.83 (1.15) & 5.50 (1.42)
& 1.04 & .594 & .03
& None \\
\bottomrule
\end{tabular}
\end{table*}

\begin{figure}[!t]
  \centering
  \includegraphics[
    width=\columnwidth,
    keepaspectratio
  ]{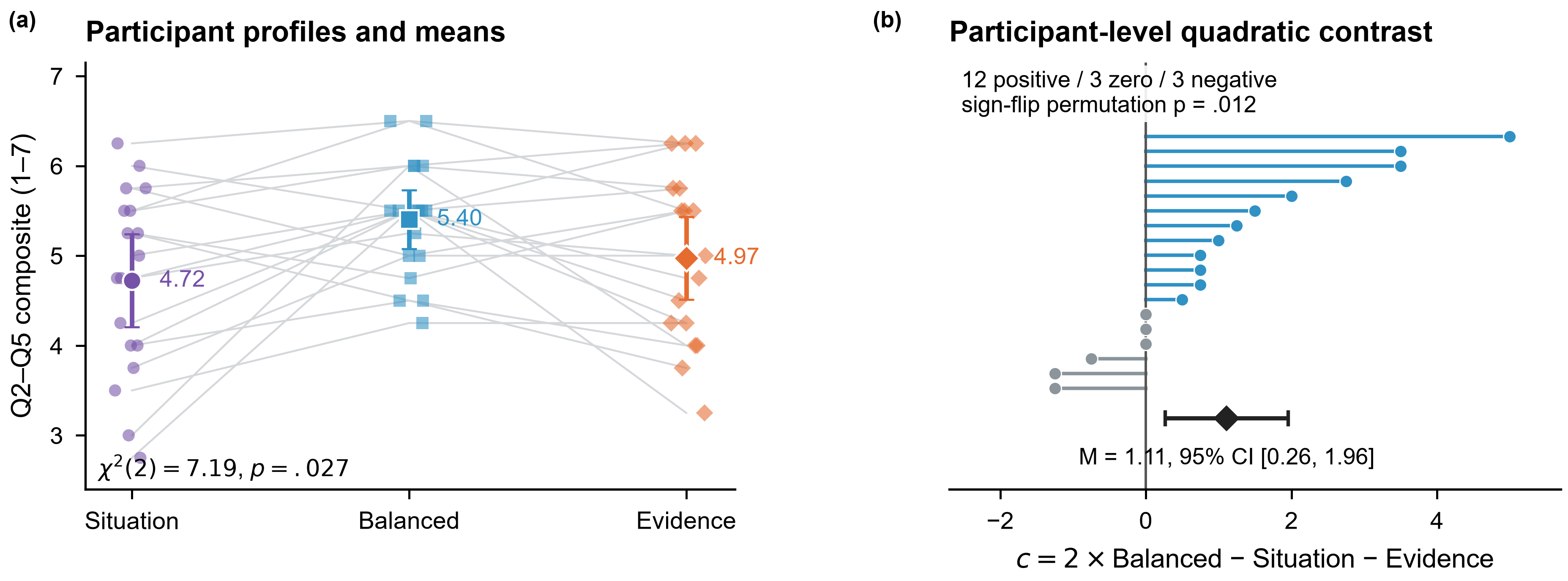}
  \caption{Exploratory cross-dimensional experience composite for Q2--Q5.
  \textbf{(a)} Participant scores, condition means, and 95\% confidence intervals under the three conditions.
  \textbf{(b)} Participant-level quadratic contrasts. Positive values indicate that the balanced condition exceeded the mean of the two extreme conditions; the diamond and error bar show the mean contrast and its 95\% confidence interval.}
  \Description{Two horizontally arranged plots of the exploratory Q2--Q5 composite. The left panel shows participant scores and condition means. The right panel shows participant-level quadratic contrasts and their mean confidence interval.}
  \label{fig:q2q5-composite}
\end{figure}

All four items used the same seven-point scale and scoring direction. Their internal consistency was $\alpha=.729$, with corrected item--total correlations of $.498$--$.532$. We therefore used each participant's mean Q2--Q5 score under each condition as an exploratory cross-dimensional experience composite, while treating the individual-item results as a descriptive decomposition of that composite. Mean composite scores were $4.72$, $5.40$, and $4.97$ in the situation-dominant, balanced, and evidence-dominant conditions, respectively. A Friedman test showed an omnibus difference among conditions, $\chi^2(2)=7.19$, $p=.027$, Kendall's $W=.20$ (Figure~\ref{fig:q2q5-composite}(a)). None of the three pairwise comparisons for the composite passed Holm correction: the balanced minus situation-dominant mean difference was $0.68$ ($p_{\mathrm{Holm}}=.057$), the balanced minus evidence-dominant difference was $0.43$ ($p_{\mathrm{Holm}}=.116$), and the comparison between the two extreme conditions yielded $p_{\mathrm{Holm}}=.218$.

Because all four items had their highest means in the balanced condition, we further constructed an exploratory quadratic contrast from each participant's composite scores:
$c_i=2\times\mathrm{balanced}_i-\mathrm{situation}_i-\mathrm{evidence}_i$.
A value of $c_i>0$ indicates that the participant rated the balanced condition above the mean of the two extreme conditions. Of the 18 participants, 12 had positive contrast values, three had values of zero, and three had negative values. The mean contrast was $M_c=1.11$, 95\% CI $[.26,1.96]$, equivalent to the balanced condition exceeding the mean of the two extreme conditions by $0.56$ scale points. An exact sign-flipping permutation test showed that the contrast was significant, $p_{\mathrm{perm}}=.012$, $d_z=0.65$ (Figure~\ref{fig:q2q5-composite}(b)); a Wilcoxon test produced a consistent result ($p=.021$). Permutation-test results across 18 leave-one-out analyses ranged from $.005$ to $.025$, indicating that the result was not driven by any single participant.

\begin{figure}[!t]
  \centering
  \includegraphics[
    width=0.82\columnwidth,
    keepaspectratio
  ]{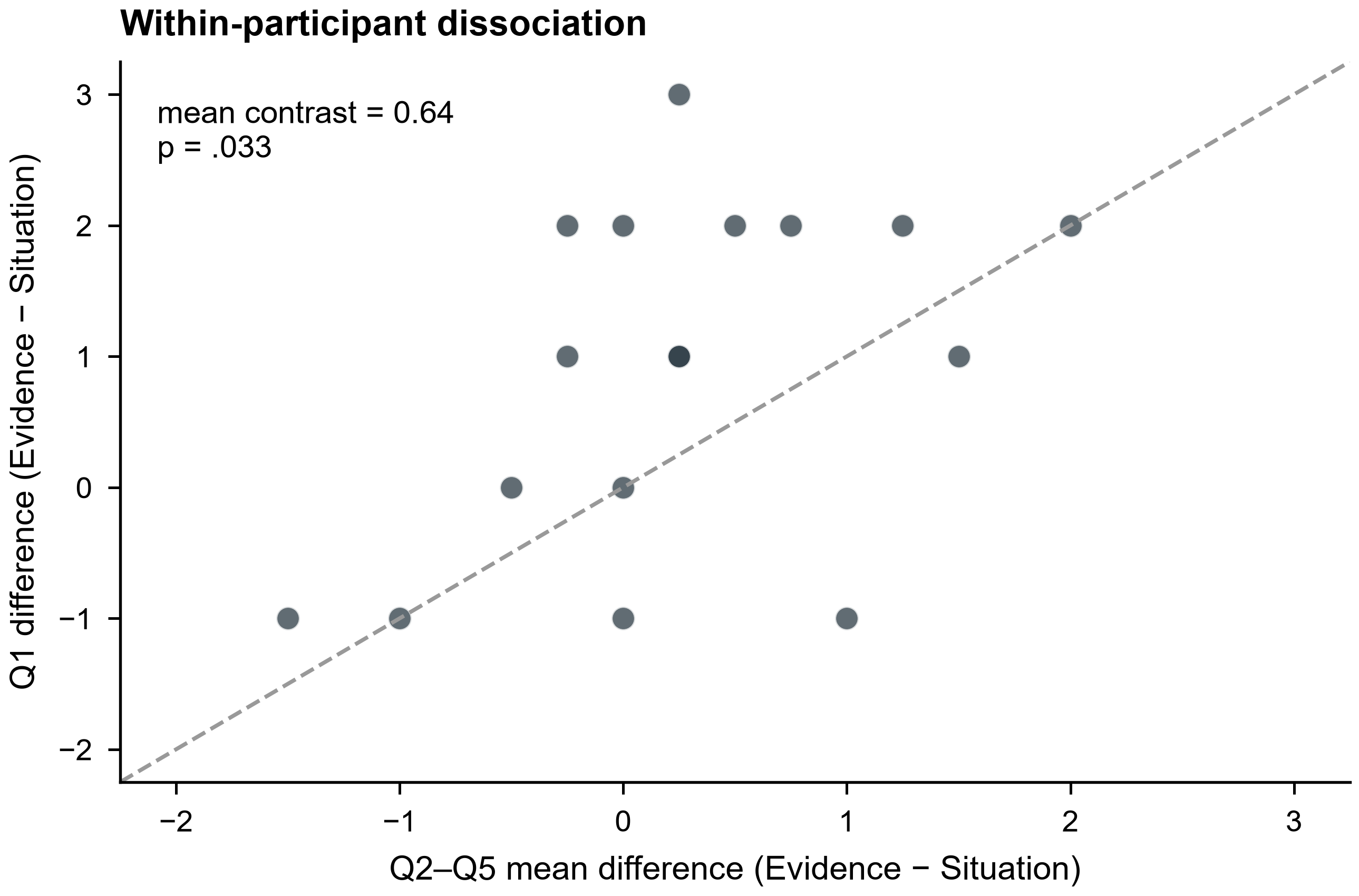}
  \caption{Dissociation between place relevance and the remaining experiential outcomes. Each point represents one participant. The horizontal axis shows the mean evidence-dominant versus situation-dominant difference across Q2--Q5, and the vertical axis shows the corresponding difference on Q1. The dashed identity line represents equal changes in Q1 and Q2--Q5. Points above the line indicate a stronger condition effect on Q1.}
  
  \Description{Participant-level dissociation contrasts between the condition effect on Q1 and its average effect on Q2--Q5. Most contrasts are positive, and the mean confidence interval is above zero.}
  \label{fig:dissociation}
\end{figure}

To test whether the condition effect on Q1 differed from the other four outcomes, we compared each participant's evidence-dominant minus situation-dominant difference on Q1 with the mean of the corresponding differences on Q2--Q5. The mean dissociation contrast was $0.64$, 95\% CI $[.11,1.15]$, $z=2.13$, $p=.033$, rank-biserial $r=.63$, indicating that the evidence-dominant versus situation-dominant difference was stronger for Q1 than for the average of the other four experiential dimensions (Figure~\ref{fig:dissociation}).

\begin{figure}[!t]
  \centering
  \includegraphics[
    width=0.82\columnwidth,
    keepaspectratio
  ]{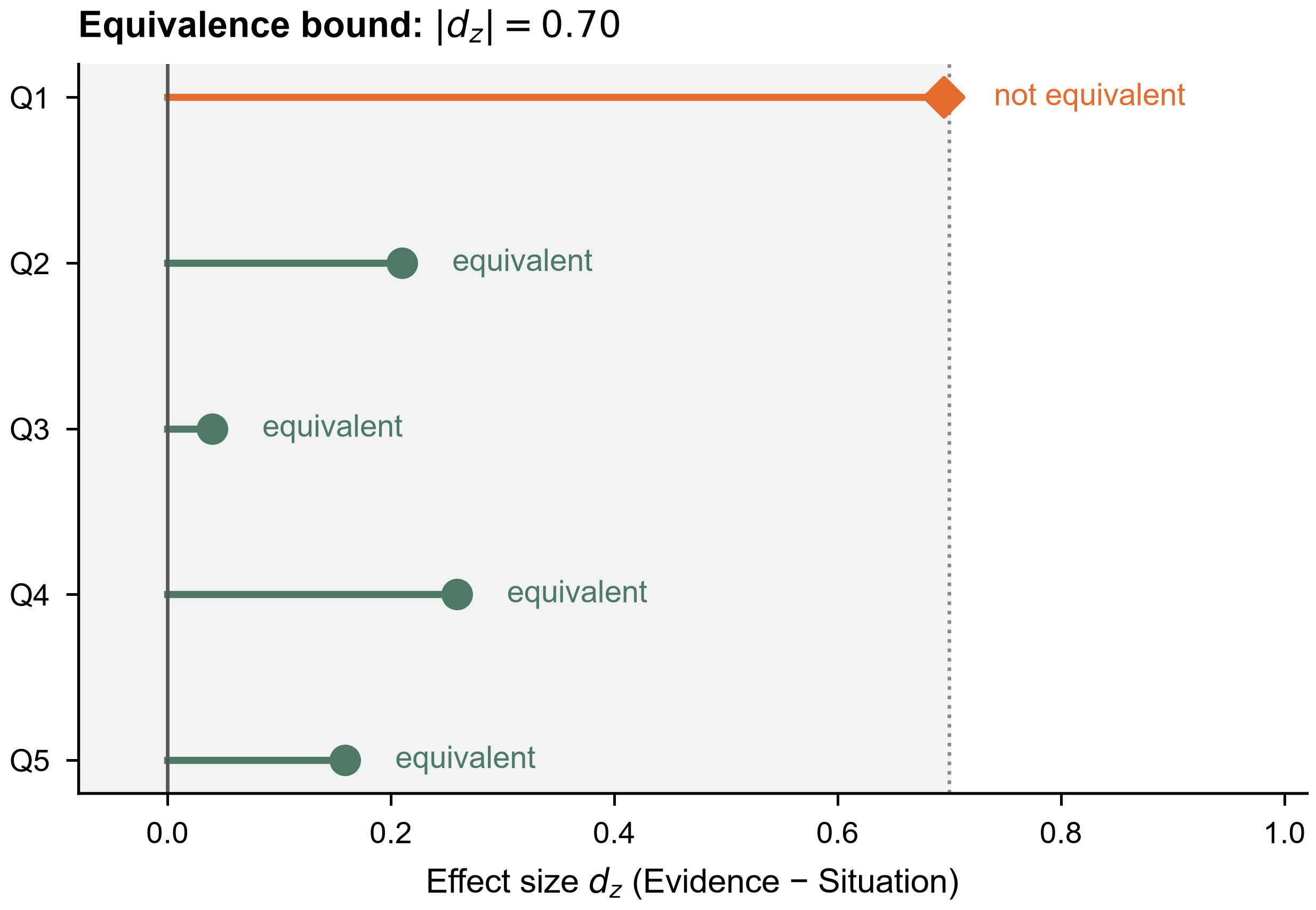}
  \caption{Equivalence tests for the evidence-dominant versus situation-dominant differences on the five experiential outcomes. Points show standardized paired mean differences ($d_z$). The shaded area represents the prespecified equivalence region bounded by $d_z=-0.70$ and $d_z=0.70$. Differences on Q2--Q5 were statistically equivalent within this bound, whereas the Q1 difference was not.}
  \Description{Standardized evidence-dominant versus situation-dominant differences for Q1--Q5 relative to the equivalence bounds of minus 0.70 and 0.70. Q2--Q5 are classified as equivalent, while Q1 is not equivalent.}
  \label{fig:equivalence}
\end{figure}

Paired-sample equivalence tests further showed that the evidence-dominant versus situation-dominant differences on Q2--Q5 were statistically equivalent within the prespecified $d_z=0.70$ bound: Q2, $p_{\mathrm{TOST}}=.027$; Q3, $p_{\mathrm{TOST}}=.006$; Q4, $p_{\mathrm{TOST}}=.039$; and Q5, $p_{\mathrm{TOST}}=.017$. The Q1 difference was not equivalent within this bound, $p_{\mathrm{TOST}}=.492$ (Figure~\ref{fig:equivalence}).

Taken together, the HGR manipulation produced a dissociation across experiential dimensions rather than a uniform response pattern. Mean Q1 ratings increased successively with HGR, with a significant difference between the two extreme conditions. By contrast, the Q2--Q5 composite reached its highest value in the balanced condition, and its quadratic contrast showed that the balanced condition exceeded the mean of the two extremes. Increasing historical grounding therefore strengthened the perceived association between narrative content and place without improving all experiential dimensions in parallel; different experiential goals may require different source-configuration strategies. Higher historical grounding was more favorable for perceived content--place relevance, whereas the combined pattern of historical understanding, scene integration, information fit, and intention to explore further favored a balanced configuration of historical evidence and situated information.

\subsection{Robustness Analysis: Narrative Length}
\label{sec:results:robustness}

\begin{figure}[!t]
  \centering
  \includegraphics[
    width=\columnwidth,
    keepaspectratio
  ]{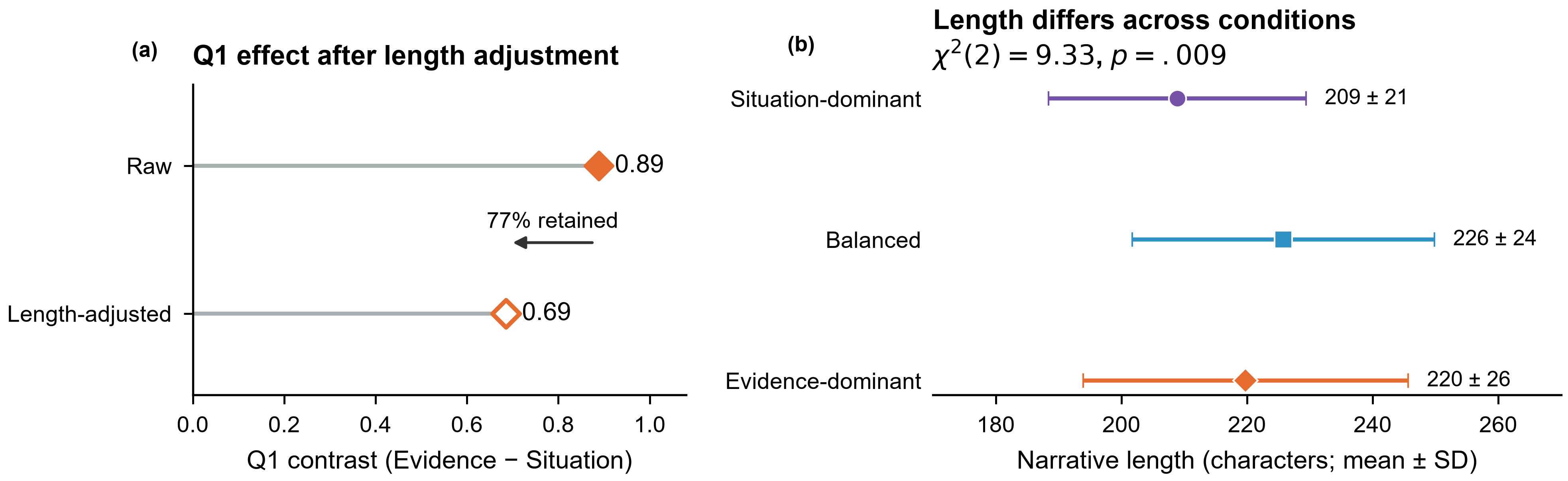}
  \caption{Robustness analysis of narrative length.
  \textbf{(a)} In the complete three-condition data from all 18 participants, the evidence-dominant versus situation-dominant mean difference in Q1 content--place relevance was $0.89$ before controlling for character length and $0.69$ after adjustment, retaining approximately $77\%$ of the original difference.
  \textbf{(b)} Narrative character length differed overall among the three historical grounding conditions, Friedman $\chi^2(2)=9.33$, $p=.009$; points and error bars show condition means and standard deviations.}
  \Description{Two-panel robustness analysis of narrative length. Panel a compares the evidence-dominant versus situation-dominant Q1 contrast before and after adjustment for character length, showing that approximately 77 percent of the original contrast remains. Panel b shows mean narrative character length and standard deviation for the three historical grounding conditions.}
  \label{fig:robustness-length}
\end{figure}

Although narratives in all three conditions contained nine sentences and similar numbers of valid information units per segment, their character lengths differed across conditions, Friedman $\chi^2(2)=9.33$, $p=.009$ (Figure~\ref{fig:robustness-length}(b)). We therefore examined the extent to which narrative length could explain the evidence-dominant versus situation-dominant difference on Q1.

After adjusting for character length using the complete three-condition data from all 18 participants, the unadjusted Q1 mean difference between the evidence-dominant and situation-dominant conditions decreased from $0.89$ to $0.69$ scale points, retaining approximately $77\%$ of the original difference (Figure~\ref{fig:robustness-length}(a)). Character length therefore explains part of the condition difference, but not most of the observed evidence-dominant versus situation-dominant difference on Q1. Text length remains an alternative explanation that must be considered, but differences in character length do not primarily drive the reported condition effect on content--place relevance.

\subsection{Post-Study Retrospective Condition Comparisons}
\label{sec:results:final}

The segment-level seven-point ratings recorded participants' independent evaluation of the current narrative at the end of each walk. As a complement, after completing all three conditions, participants used a retrospective comparison questionnaire to select: (1) the narrative they preferred overall; (2) the narrative with the most appropriate amount of historical information; and (3) the narrative that created the strongest sense of connection to the current place. The former provides absolute ratings under each condition, whereas the latter requires a direct within-subject comparison after all conditions have been experienced.

The balanced narrative received the most selections for overall preference (9/18) and appropriateness of historical information amount (10/18), whereas the evidence-dominant narrative received the most selections for sense of connection to place (9/18). None of the three choice distributions differed significantly from a distribution in which the conditions were equally likely to be selected (exact multinomial tests: $p=.419$, $.180$, and $.243$, respectively).

At the participant level, 10/18 selected the same condition for overall preference and information fit, whereas only 5/18 selected the same condition for information fit and connection to place. Considered alongside the post-segment ratings and Q2--Q5 composite, the retrospective comparisons show the same differentiation in outcomes. The balanced condition was more often associated with overall preference and information fit, while the evidence-dominant condition was more often associated with connection to place. The most preferred narrative configuration was therefore not identical to the configuration that most strongly enhanced connection to place.

\subsection{Participant Reflections Explaining the Quantitative Patterns}
\label{sec:results:qual}

All 18 participants contributed interview material, which was transcribed verbatim. Table~\ref{tab:qual} summarizes reflection themes inductively derived from the transcripts, presenting participants' main perceptions of the narrative experience and their relationship to the quantitative results. One analyst generated the themes through open-ended induction, and the material is used primarily to provide supplementary explanations for experiential differences observed in the experiment.

The analysis identified 13 themes. Table~\ref{tab:qual} lists the five themes directly related to the main quantitative results and their design implications. The remaining feedback concerning speech quality, visual richness, interaction methods, and social or commercial functions primarily reflected participants' broader expectations of intelligent interactive systems rather than experiential dimensions manipulated through the source composition of narrative content.

\begin{table*}[t]
\centering
\caption{Participant reflection themes directly related to the main quantitative results. Counts indicate the number of participants for whom a corresponding reference could be located in the transcript; they are not estimates of population prevalence.}
\label{tab:qual}
\small
\begin{tabular}{p{0.22\textwidth}cp{0.27\textwidth}p{0.37\textwidth}}
\toprule
Theme & Count & Corresponding quantitative pattern & Analyst's summary of reflections \\
\midrule

Differences in prior knowledge
& 8
& Between-participant variation under all conditions
& Participants differed in local familiarity and disciplinary background and suggested adapting information organization to the prior knowledge of different users. \\

Spatial guidance and route organization
& 4
& Q3 and Q5 had their highest means under the balanced condition
& Participants wanted clearer spatial guidance and route structure and felt that walking narratives should be more closely organized around movement. \\

Narrative anchors and the current location
& 3
& Q1 increased with HGR
& Participants wanted specific narrative objects to correspond directly to their current location rather than merely describing generic features of the scene. \\

Historical authenticity and connection to place
& 3
& Q1 increased with HGR; evidence-dominant received the most selections for connection to place
& Participants were concerned with the credibility of archival materials and whether historical figures, events, and temporal information could be clearly associated with the current place. \\

Timeline coherence and narrative discontinuity
& 3
& Q2 had its highest mean under the balanced condition
& Participants felt that some narratives shifted abruptly among historical details or jumped across time, increasing the burden of comprehension. \\

\bottomrule
\end{tabular}
\end{table*}

Participant reflections provide additional context for interpreting the two quantitative patterns. Feedback related to Q1 focused primarily on whether specific narrative objects corresponded to the current location and on the authenticity and place attribution of historical materials. Content--place relevance may therefore depend not only on the amount of scene description, but also on whether historical information clearly bounds the narrative to the current place. Feedback related to the balanced condition more often concerned route organization, timeline coherence, and comprehension burden, suggesting that the overall experience also depends on whether historical evidence and the mobile context are organized into a coherent narrative. Differences in prior knowledge further indicate that participants may receive the same information configuration differently. Because a single analyst derived these themes, the material is used to propose possible explanatory mechanisms rather than to validate condition effects.


\section{Discussion}
\label{sec:discussion}

\subsection{Differentiated Effects of HGR on Situated Narrative Experience}
\label{sec:disc-results}

We first examined whether HGR, as a parameter of narrative source composition, could produce perceptible experiential differences. Achieved HGR separated clearly across the three conditions as intended, and participants' ratings of relevance between the narrative and the current place increased with historical grounding. The evidence-dominant condition scored significantly higher than the situation-dominant condition on Q1, indicating that increasing the proportion of historical evidence in the final text can strengthen participants' perception that a situated historical narrative genuinely belongs to the current place. The source-composition results further show that this change primarily involved archive-supported information units progressively replacing scene-only and interpretive units, while the numbers of information units and sentences remained similar across conditions. Participants therefore responded not only to how much content a narrative contained, but also to the sources supporting that content and the share of the final narrative occupied by historical material. Text length remains a partial alternative explanation, but the robustness analysis retained approximately 77\% of the original Q1 difference between the evidence-dominant and situation-dominant conditions after adjustment for character length (Section~\ref{sec:results:robustness}). Character length alone is therefore insufficient to explain the result.

This finding supports our basic expectation that historical grounding can strengthen perceived place relevance, but the other experiential dimensions followed a different pattern. Q2--Q5 did not increase with HGR. Instead, all four had their highest means in the balanced condition, as did the exploratory cross-dimensional composite, and the quadratic contrast between the balanced condition and the mean of the two extremes was positive. The retrospective comparisons showed a similar differentiation: overall preference and appropriateness of historical information more often favored the balanced condition, whereas connection to place more often favored the evidence-dominant condition. These results are therefore better understood as a \emph{dissociation} among experiential outcomes than as a uniform response curve applying to every experiential goal. Increasing HGR more clearly strengthened the perceived association between content and place, but did not improve historical understanding, scene integration, information fit, and intention to explore further in parallel.

This dissociation also limits how the balanced condition should be interpreted. The Q2--Q5 composite provides exploratory evidence for a better cross-dimensional experience at an intermediate level of historical grounding, but it does not establish a universally optimal HGR value or support a strict inverted-U relationship. The three conditions represent discrete information-configuration strategies rather than systematic sampling of a continuous HGR space, and achieved HGR also varied within conditions. The appropriate conclusion is therefore not that one fixed proportion outperforms all others, but that different experiential goals may impose different information-configuration requirements. Higher historical grounding appears advantageous for tasks emphasizing the relationship between historical facts and a specific place. For an integrated experience that simultaneously requires understanding, scene integration, information fit, and continued exploration, the results motivate further study of configurations that balance historical evidence with situated and interpretive content.

Scene integration (Q3) warrants further discussion. The most direct expectation is that reducing HGR and increasing the proportion of scene-supported content should make a narrative easier to integrate with the environment in front of the pedestrian, but we did not observe this monotonic relationship. Historical evidence and situated information may not form a simple trade-off. Specific people, events, dates, and spatial objects in historical materials may provide identifiable references through which users understand the visible environment, giving situated information an object around which it can be organized and interpreted. Under such conditions, historical grounding need not compete with scene integration and may instead provide structure for scene understanding. At the same time, the runtime records show that some generation events lacked a real-time scene description and that character length still differed across conditions. The present data therefore cannot isolate the independent contributions of source composition, completeness of situated input, and textual expression to Q3.

Exploratory text analyses provide additional clues. After controlling for text length, explicit date information remained associated with Q1, whereas the relationship between proper-name density and Q1 weakened substantially. Atmospheric sentences lacking a specific object or temporal reference were negatively associated with scene integration. Although these analyses do not establish a particular linguistic mechanism, they suggest that historical grounding may do more than increase historical content: it may provide narratives with objects that can be located, situated in time, and identified. For situated storytelling, what matters may not be simply adding place names or scene descriptions, but establishing an intelligible relationship among historical objects, spatial objects, and the user's current location.

Participant reflections further support this interpretation. Feedback corresponding to Q1 focused on whether narrative anchors genuinely matched the current location, whether the historical materials were credible, and whether people, events, and temporal information could be clearly attributed to the place. Feedback corresponding to the stronger integrated experience under the balanced condition more often concerned route organization, timeline coherence, and comprehension burden while walking. Differences in participants' prior knowledge also show that users do not receive the same source configuration in identical ways. Together with the quantitative results, this feedback indicates that source composition affects not a single notion of content quality, but how historical materials, the in-situ environment, and users' prior knowledge are organized during a particular walk.

\subsection{From Source Composition to Interface Policy: Design Implications of HGR}
\label{sec:disc-implications}

Our primary contribution to generative intelligent interfaces is not a new general-purpose quality metric, but the explicit treatment of the \emph{relative source contribution} to a final output as an interface variable that can be measured, manipulated, and connected to user experience. Traditional context-aware systems primarily address when to acquire context, what information to retrieve, and how to filter content according to location and environment. Once a generative model also organizes and expresses content, the interface must additionally determine the proportions in which retrieved sources enter the output users actually receive. Prior work can explicitly model location, the visual environment, and other contextual signals as generation conditions~\cite{han2025multimodal,liu2025gensors}. Our findings further show that, after these sources enter the generation process, their relative contributions to the final text can themselves become a researchable interface policy.

HGR operationalizes this principle for urban historical narratives. It transforms the realized contribution of historical archives to a delivered narrative from an implicit post-generation outcome into a parameter for which a target can be set during interface design, measured in the output, and evaluated through user studies. This also distinguishes HGR from retrieval-quality and fact-checking metrics. A system may retrieve equally relevant historical materials, yet whether those materials actually enter the final narrative, the share they occupy, and how they are organized with situated and interpretive content can still produce different user experiences. Retrieval relevance therefore cannot fully proxy the final interface outcome, consistent with generation-evaluation frameworks that treat context relevance and answer relevance separately~\cite{es2024ragas,saadfalcon2024ares}.

More importantly, the observed dissociation indicates that source allocation should not be designed as a monotonic quality dial. Higher HGR better strengthened the perceived association between content and a specific place, whereas the cross-dimensional experience more strongly favored the balanced configuration. Interface policy should therefore begin by identifying the experience to optimize rather than assuming that more historical evidence necessarily improves the overall experience. A system can increase historical grounding when it needs to emphasize that a historical event is tied to the current location. When users are walking continuously and must both understand the environment and maintain narrative coherence, more expressive space may be needed for situated cues and interpretive connections. HGR is thus better treated as an interface policy configured according to a design goal than as a quality score for which higher is always better. Stronger coupling between a narrative and a physical place does not in itself imply parallel improvements across all experiential dimensions~\cite{karapanos2012locality}.

This view points toward dynamic source allocation. The present experiment used a relatively stable information configuration within each narrative segment, but location, visible objects, user attention, and prior knowledge change continuously during real-world walking. Future situated intelligent interfaces need not assign one fixed HGR to an entire route; they could adjust source composition dynamically across locations and interaction stages. For example, a system might increase historical grounding upon entering a location associated with a well-defined historical event or architectural remain, and increase situated description and interpretive connection when moving between locations or when attentional demands are high. Users could also actively change the source configuration of subsequent narratives by asking follow-up questions, requesting more historical material, or asking the system to return attention to the visible environment. Proactive context awareness and information selection are beginning to appear in research on generative interfaces~\cite{cai2025aiget}; our results indicate that source contribution itself can become a control target for such adaptive strategies.

Participant feedback on speech quality, visual richness, interaction methods, and social or commercial functions also highlights that HGR captures one dimension of content organization in an intelligent narrative interface, not the complete interaction experience. Participants wanted to continue interacting with the system in different ways: some wanted richer visual or spoken presentation, some emphasized route and action feedback, and others expected further question answering, social interaction, or functional support. Situated storytelling therefore need not be designed as the endpoint of an experience. A generated local narrative can first establish shared context between a user and a place, then serve as an entry point for questions, navigation, historical exploration, or other interactions. In such systems, HGR can control how the content entering that interaction is distributed among historical evidence, situated information, and interpretation, while other interface policies determine conversational depth, presentation modality, and social interaction. Designing generative intelligent interfaces should therefore not seek a single parameter that explains the entire experience, but coordinate source configuration with interaction method, presentation form, and user control.

This framework also has direct implications for digital cultural heritage and urban historical experiences. Traditional heritage narratives rely on curators, guide texts, or pre-edited content to determine which historical materials enter public experience. Generative systems partially delegate this choice to a real-time generation process. When the way different sources enter the final narrative becomes invisible, heritage institutions cannot easily determine the extent to which a system relies on traceable historical sources, in-situ observation, or connective language produced by the model. HGR cannot establish that a narrative is accurate or replace expert review of historical materials, but it provides a way to record and inspect source composition. Combining this measurement with user-visible source cues, steering controls, or audit mechanisms could make the organization of generated content more inspectable~\cite{hoque2024hallmark}. More generally, our contribution is to transform source allocation from an implicit outcome within model generation into an interface-level problem that can be designed, evaluated, and audited.

\subsection{Research Boundaries and Future Directions}
\label{sec:limitations}

Our goal was to establish whether source composition can serve as an operational design variable in situated generative narratives and to explore the experiential outcomes associated with different configurations, not to identify an optimal HGR that generalizes across places, systems, and users. The experiment covered one walking route in one historical urban district, used a single model configuration, and recruited 18 participants for within-subject comparisons. The present results therefore apply first to the urban historical walking context studied here. Future research should replicate the manipulation in places with different densities of historical information, spatial forms, and user populations, and compare whether source configurations produce similar differentiated effects across generative models and narrative forms.

The experiment used three discrete conditions to form low, medium, and high regions of historical grounding. This enabled clear within-subject comparisons but cannot identify the response function across a continuous HGR space. Achieved HGR also varied within conditions, so target configurations in prompts cannot be equated directly with realized values. Future work could specify more target grounding levels, sample HGR more densely with larger samples, and verify the manipulation using measured values from delivered texts. Such studies could test whether the apparent advantage of an intermediate level is stable and whether different experiential goals correspond to different appropriate ranges.

Text length is another important interpretive boundary. The generator controlled the numbers of sentences and information units, but did not fully equalize character length. The robustness analysis retained approximately 77\% of the original Q1 difference between the evidence-dominant and situation-dominant conditions after controlling for character length. Text length therefore cannot explain the main result, but it can explain part of it. The present study cannot completely separate the effects of source composition and expression length. Future experiments should constrain the numbers of information units and sentences together with text length during generation, or prespecify length as a covariate in the analysis plan, to more strictly distinguish the experiential effect of source configuration itself.

The measurement scope of Q1 also requires refinement. The current single item measures participants' perceived relevance between narrative content and the current place, but this judgment may combine place attribution, referential specificity, and credibility of historical materials. Exploratory text results suggest that historical content with explicit temporal references may produce this specificity more effectively than simply adding place names. Future research should separate place relevance, referential specificity, historical credibility, and situatedness into distinct constructs and use multi-item measures to determine which place-related experience HGR changes first.

Incomplete real-time scene input primarily limits our explanation of the scene-integration mechanism. Some generation events did not obtain an available real-time scene description and therefore could not execute the planned scene-aware generation process in full. This issue did not eliminate the overall manipulation difference among the three HGR conditions, but makes it difficult to determine why the low-HGR condition did not show the expected advantage in scene integration. Future studies should ensure consistent-quality situated input for every experimental segment and distinguish whether situated information is retrieved, whether it enters the generation context, and whether it ultimately appears in the narrative received by the user.

The interaction form also has clear boundaries. All three conditions primarily used continuous audio narratives, and participants did not alter the information configuration within an individual segment through sustained follow-up questions or active requests. The study therefore examines system-directed source allocation at generation time rather than a full conversational process in which users and the system negotiate narrative depth. Participants' expectations of additional interaction methods suggest that the effect of preset HGR may change once users can actively request more historical material, ask for explanations of visible objects, or redirect the narrative. Future work could extend HGR from a parameter for a single generation into a dynamic state jointly regulated by user behavior and context, enabling study of how system control and active user control together determine source configuration.

Finally, HGR should be understood as a measurement parameter that depends on explicit source-classification and information-unit segmentation rules, not as a universal score that can be compared independently of its annotation protocol. We annotated all 54 narrative segments with a model and examined segment-level HGR consistency through human verification of 11 segments, providing basic measurement support for the condition comparisons in this experiment. Other systems, however, may use different source categories, granularities, and evidentiary standards. Two narratives with the same HGR may still differ substantially in the importance of their historical information, evidence quality, and narrative organization. Cross-system comparisons of HGR must therefore report both information-unit segmentation and source-classification rules. Building larger human-annotated datasets with information units aligned across annotators would help assess the measurement stability of HGR across narrative types and source structures.

These boundaries do not alter the central observation established by this study: when the generation process and narrative structure remain broadly consistent, changing source composition in the final output can systematically change parts of the user experience, and the effect differs across experiential goals. The question for subsequent research is not how to find a universally maximal HGR, but which source-configuration strategy should be used for a particular place, task, user state, and stage of interaction.

\section{Conclusion}
\label{sec:conclusion}

Situated generative narratives jointly organize historical archives, the in-situ environment, and interpretive content produced by generative models. Yet the proportions in which these sources enter the narrative users actually receive have rarely been represented and measured as an interface design variable. We make this source-allocation process explicit by introducing the Historical Grounding Ratio (HGR) as a design parameter for situated generative narratives. HGR measures the proportion of claim-bearing information units in the delivered text that are supported by historical archives. It concerns not how much historical material the system retrieves, but the extent to which that material actually enters the narrative experienced by users.

The within-subject field study showed that HGR can be effectively manipulated through different source configurations, producing clearly separated achieved grounding levels in the generated texts. As HGR increased, participants perceived a stronger relationship between narrative content and the specific place, while the other experiential dimensions did not follow the same monotonic pattern. Further cross-dimensional analysis provided exploratory evidence that the balanced configuration offered an advantage in the combined pattern of historical understanding, scene integration, information fit, and intention to explore further. The present study, however, cannot establish a universally optimal HGR range.

The study therefore reveals a dissociation in the effects of historical grounding across experiential outcomes: increasing HGR strengthens participants' perceived association between narrative content and a specific place, but does not improve all experiential dimensions in parallel. Higher historical grounding is better suited to emphasizing the relationship between narrative content and a particular place, whereas a more integrated experience may require a different degree of balance among historical evidence, situated information, and interpretive content. HGR should therefore not be understood as a monotonic generation-quality score, but as an interface-policy parameter that must be configured and validated according to a specific experiential goal.

More generally, we transform source contribution in generative interfaces from an implicit outcome of generation into a manipulable and measurable design object. The study also shows that information quotas in prompts cannot be treated directly as the grounding level ultimately achieved. A generative model's implementation of source requirements can still deviate in the final expression. Any design that depends on a particular HGR should therefore measure its achieved value in the delivered text rather than infer it solely from prompts or generation settings. For situated historical narratives, the central issue is no longer simply how to add more historical information, but how historical evidence, situated information, and interpretive content jointly enter the final narrative as an intelligent-interface mechanism that can be designed, measured, and adjusted according to experiential goals.

\section*{Ethics Statement}

This study received institutional ethics approval before data collection. All participants were adults, participated voluntarily, and provided electronic informed consent before data collection. Participants were informed that the generated narratives might contain errors and could withdraw from the study. No directly identifying participant information is reported.

\section*{GenAI Usage Disclosure}

Generative AI was used both as the object of study in the GeoDrama system and as research assistance. The evaluated GeoDrama system used Qwen/Qwen3-8B for narrative generation. Claude and OpenAI Codex were used to assist with language revision, analysis scripting, code debugging, and figure production. All study-design decisions, data collection procedures, statistical results, interpretations, citations, and final manuscript content were reviewed and verified by the authors. The authors take full responsibility for the accuracy and integrity of the submitted work.

\bibliographystyle{ACM-Reference-Format}
\bibliography{refs}
\end{document}